\documentclass[12pt,preprint]{aastex}
\begin{document}

\parindent=1.0cm

\title{CLUSTER GLIMPSES with RAVEN: AO-CORRECTED NEAR and MID-INFRARED IMAGES OF 
GLIMPSE C01 and GLIMPSE C02
\altaffilmark{1,}\altaffilmark{2}}

\author{T. J. Davidge, D. R. Andersen}

\affil{Dominion Astrophysical Observatory,
\\National Research Council of Canada, 5071 West Saanich Road,
\\Victoria, BC Canada V9E 2E7\\tim.davidge@nrc.ca, david.andersen@nrc.ca}

\author{O. Lardi\`{e}re, C. Bradley, C. Blain}

\affil{Department of Mechanical Engineering, University of Victoria, 
\\Victoria, BC Canada V8W 3P2\\lardiere@uvic.ca, cbr@uvic.ca, celia.blain@gmail.com}

\author{S. Oya, H. Terada, Y. Hayano}

\affil{Subaru Telescope, National Optical Observatory of Japan\\Hilo, HI USA 96720 \\
oya@subaru.naoj.org, terado@naoj.org, hayano@naoj.org}

\author{M. Lamb}

\affil{Department of Physics \& Astronomy, University of Victoria, 
\\Victoria, BC Canada V8W 3P2\\masen@uvic.ca}

\author{M. Akiyama, Y. H. Ono, \& G. Suzuki}

\affil{Astronomical Institute, Tohoku University\\6--3 Aramaki, Aoba-ku, Sendai\\
Japan 980-8578\\ akiyama@astr.tohoku.ac.jp, yo-2007@astr.tohoku.ac.jp, g.suzuki@astr.tohoku.ac.jp}

\altaffiltext{1}{Based on data obtained at Subaru Telescope, which is operated by 
the National Optical Observatory of Japan.}

\altaffiltext{2}{This research has made use of the NASA/IPAC Infrared Science Archive,
which is operated by the Jet Propulsion Laboratory, California Institute of Technology,
under contract with the National Aeronautics and Space Administration.}

\begin{abstract}

	We discuss images of the star clusters 
GLIMPSE C01 (GC01) and GLIMPSE C02 (GC02) that were recorded with the 
Subaru IRCS. Distortions in the wavefront were corrected 
with the RAVEN adaptive optics (AO) science demonstrator, allowing individual stars 
in the central regions of both clusters -- where the fractional contamination from 
non-cluster objects is lowest -- to be imaged. In addition to $J, H,$ and $K'$ 
images, both clusters were observed through a narrow-band filter centered near 
3.05$\mu$m; GC01 was also observed through two other narrow-band filters that 
sample longer wavelengths. Stars in the narrow-band images have a FWHM that is close 
to the telescope diffraction limit, demonstrating that open loop AO systems like RAVEN 
can deliver exceptional image quality. The near-infrared color magnitude 
diagram of GC01 is smeared by non-uniform extinction with a $1 \sigma$ 
dispersion $\Delta A_K = \pm 0.13$ magnitudes. Spatial variations in A$_K$ are not 
related in a systematic way to location in the field. The Red Clump is identified in 
the $K$ luminosity function (LF) of GC01, and a distance modulus of 13.6 is found. 
The $K$ LF of GC01 is consistent with a system that is dominated by stars with an 
age $> 1$ Gyr. As for GC02, the $K$ LF is flat for $K > 16$, and the absence of a 
sub-giant branch argues against an old age if the cluster is at a distance of $\sim 
7$ kpc. Archival SPITZER [3.6] and [4.5] images of the clusters are also examined, and 
the red giant branch-tip is identified. It is demonstrated in the Appendix
that the [3.6] surface brightness profiles of both clusters 
can be traced out to radii of at least 100 arcsec.

\end{abstract}


\section{INTRODUCTION}

	Star clusters are fundamental astrophysical calibrators, providing 
information that can be used to constrain the evolution of stars, stellar systems, 
and galaxies. From a technical perspective, they are also ideal targets for 
characterizing the performance of adaptive optics (AO) systems, as the image 
quality and its variation with location across the science field can be assessed 
in a straight-forward manner from images of richly populated stellar fields. In 
this paper we investigate the stellar contents of two star clusters 
at low Galactic latitude and demonstrate the performance of the RAVEN 
multi-object AO system.

	Glimpse C01 (GC01) is a massive (log(M$_{\odot}) \sim 5$) cluster 
that was identified as part of the GLIMPSE (Galactic Legacy 
Infrared Mid-Plane Survey Extraordinaire; Benjamin et al. 2003) survey. 
The radial velocity of GC01 is consistent with it 
belonging to the Galactic disk, although there is a 10\% probability that 
a halo object would have a similar radial velocity (Davies et al. 2011). 
Dust and contamination from non-cluster sources are major 
obstacles for efforts to probe the stellar content of GC01. 
Near-infrared (NIR) images reveal dust lanes in and around GC01 
(Ivanov et al. 2005), and a bright emission feature cuts across [5.8] and 
[8.0] SPITZER images of the cluster (Kobulnicky et al. 2005). The location of stars 
in the $(J-H, H-K)$ two-color diagram (TCD) indicates that A$_V$ ranges between 12 
and 18, with no systematic dependence on location (Kobulnicky et al. 2005). 
Such a non-uniform dust distribution will smear features in 
color-magnitude diagrams (CMDs) and luminosity functions (LFs). 

	Previous studies of GC01 have found a wide range of possible ages. 
Using a mix of GLIMPSE survey and shallow ground-based NIR images, 
Kobulnicky et al. (2005) conclude that GC01 is an old, 
massive globular cluster, located at a distance of 3.1 -- 5.2 kpc. They 
note that an old age is consistent with a lack of radio emission. 
Ivanov et al. (2005) construct a shallow CMD of sources in the central 
20 arcsec of GC01. They identify a giant branch and a red clump (RC), and 
conclude that if the former sequence is populated by old 
red giant branch (RGB) stars then [Fe/H] $\leq -1$. Ivanov 
et al. (2005) estimate a distance of $3.8 \pm 0.7$ kpc from 
the brightnesses of the RC and the tip of the giant branch. Davies et al. (2011) 
measure a central mass density for GC01 that exceeds that in globular clusters, 
but is consistent with that in dynamically unevolved young clusters, 
such as the Arches (e.g. Espinoza, Selman, \& Melnick 2009). Based on the high 
central density and other lines of evidence, Davies et al. 
(2011) suggest that GC01 has an age between 400 and 800 Myr, but also state that 
ages up to 2 Gyr are not ruled out. 

	There are hints that GC01 may be experiencing significant evolution at 
the present day, making it a potentially important laboratory for studies of cluster 
evolution, while also providing additional clues into its age. 
Mirabel (2010) discusses an x-ray source that is located along the MIR 
emission feature that slices through the cluster, and suggests that it
is either the result of a pulsar wind nebula -- possibly 
associated with a cluster member -- or emission from a bow shock that forms as 
interstellar gas associated with GC01 is stripped from the cluster. 
The location of the source and its energy output are 
consistent with the latter mechanism (Mirabal 2010). If there is an interstellar 
medium (ISM) in GC01 then it opens the possibility that a young population 
might be present. 

	GC01 has a mass $(8 \pm 3) \times 10^4$ M$_{\odot}$ 
(Davies et al. 2011). There are young clusters 
with comparable masses in the present-day Galaxy, such as Westerlund 1 
and 2 ($6 \times 10^4$ and $7 \times 10^3$ solar; Portegies Zwart, 
McMillan, \& Gieles 2010; Hur et al. 2015), and the Arches ($2 \times 10^4$ solar; 
Espinoza et al. 2009). Given the likelihood that GC01 has lost mass due to 
tidal effects and internal evolution, and so was more massive in the past, 
the existing age estimates suggest that it could be one of the most massive clusters 
to have formed in the Galaxy during the past few Gyr. GC01 is thus of 
potential importance for studies of the evolution of the Galactic disk. 

	Glimpse C02 (GC02) has not been as extensively studied as GC01, likely 
because it is the more heavily obscured of the two, with A$_V > 20$ (Kurtev et 
al. 2008). The distribution of points on the 
$(J-H, H-K)$ TCD shown in Figure 3 of Kurtev et al. (2008) indicates that there is 
substantial field star contamination within 60 arcsec of the cluster center, 
further complicating efforts to determine cluster properties.
Still, the CMD presented by Kurtev et al. (2008) includes stars as 
faint as $K \sim 16$, from which they estimate a distance of $4.6 \pm 0.7$ kpc 
based on the brightnesses of the RC and the tip of the red sequence, which they 
assume to be populated by stars evolving on the RGB. Spectroscopy of 
candidate cluster members and the slope of the red sequence suggest that 
[Fe/H] $\sim -0.3$, raising the prospect that GC02 may be one of the most metal-rich 
globular clusters in the Galaxy. 

	The existing studies of GC01 and GC02 do not sample the main 
sequence turn-off (MSTO). While reaching the MSTOs of these clusters will be difficult 
-- and it is noted later in this paper that measuring the brightness of the 
MSTO in GC01 without spectroscopic information may prove to be 
problematic due to differential reddening -- 
it is still possible to gain additional information about 
their ages based on the properties of evolved 
cluster members, such as those that are undergoing core 
Helium burning. The fractional contamination from non-cluster stars is lowest 
in the central regions of GC01 and GC02, although the high stellar density 
introduces complications due to crowding. Efforts to isolate stars 
in the crowded central regions of clusters require good 
angular resolution, and in the present paper we discuss observations 
that cover the $1 - 3.5\mu$m wavelength interval with angular resolutions between 
0.1 and 0.25 arcsec FWHM of fields near the centers of both clusters. The data were 
obtained with the Infrared Camera and Spectrograph (IRCS) on the Subaru telescope, 
with the wavefront corrected for atmospheric distortion by the RAVEN AO science 
demonstrator. These observations demonstrate that angular resolutions close to
the telescope diffraction limit can be obtained with a multi-object AO (MOAO) 
system at wavelengths near $3\mu$m. The Subaru observations are supplemented 
with archival SPITZER images of both clusters. 

	The paper is structured as follows. Details of the observations and 
the steps used to process the images are presented in Section 2. The 
CMDs, luminosity functions (LFs), and TCDs that were extracted from the NIR 
images are discussed in Section 3. A photometric analysis of 
cluster stars at wavelengths longward of $3\mu$m that utilizes 
the narrow-band IRCS images and archival [3.6] and [4.5] SPITZER images 
follows in Section 4. The paper closes with a summary and 
discussion of the results in Section 5.

\section{OBSERVATIONS \& REDUCTION}

\subsection{The Observations}

	The data were recorded at the Subaru telescope during parts of three nights 
in June and July 2015. Distortions in the wavefront were corrected using the RAVEN 
MOAO science demonstrator (Andersen et al. 
2012; Lardi\'{e}re et al. 2014), with the corrected signal 
directed to the imaging arm of the IRCS (Tokunaga et al. 1998). The IRCS $1024 \times 
1024$ ALLADIN III detector can be sampled with pixel scales of either 
0.02 arcsec/pixel or 0.052 arcsec/pixel, and both modes were employed here. 

	Core observational elements of RAVEN include 
(1) three natural guide star (NGS) wavefront sensors (WFSs) that can be 
deployed over a 3.5 arcmin diameter field, and 
(2) two science pick-offs, each of which contains an $11 \times 11$ element 
deformable mirror (DM) that corrects the wavefront at that location using 
information gleaned from the WFSs. There is also a WFS designed for use with a 
laser beacon, but this was not used for these observations. 
The light from the science pick-offs can feed 
the imaging and spectroscopic modes of the IRCS. 
Each pick-off samples a 5.5 arcsec radius field, although 
vignetting and a slight overlap of the science fields when projected onto the 
IRCS detector limits the useable field to $\sim 6 \times 9$ arcsec. RAVEN was built 
as a pathway science demonstrator with a limited budget. Future MOAO systems will 
likely include more NGSs and science pick-offs to increase the field of view 
and the order of correction, thereby better exploiting the multiplex advantage 
that can be realised with MOAO.

	RAVEN has three operating modes: MOAO, ground-layer AO (GLAO), and classical 
single conjugate AO (SCAO). Wavefront corrections for the MOAO and GLAO modes 
are applied with the system operating open loop -- i.e. the control of the DMs is 
based solely on the signal obtained from the WFSs at that moment with 
no feedback from previous corrections. The SCAO system runs 
closed-loop, in which information from past corrections is used to 
control the DMs. The observations discussed here were recorded in GLAO mode. 

	An observing log that lists filters, central wavelengths, total 
exposure times, pixel sampling, FWHM, and the dates of observation is shown in 
Table 1. The total exposure time entries in this table are the number of detector 
co-adds (i.e. the number of detector reads $\times$ the number of co-adds per read) 
$\times$ the integration time per co-add. Additional information 
about the filters can be found on the Subaru telescope website \footnote[3]
{http://www.naoj.org/Observing/Instruments/IRCS/camera/filters.html}.

\begin{table*}
\begin{center}
\begin{tabular}{clccccl}
\hline\hline
Cluster & Filter & $\lambda_{center}$  & Exposure Time & Pixel Scale & FWHM & Date Observed\\
 & & ($\mu$m) & (seconds) & (arcsec/pixel) & (arcsec) & in 2015 (UT) \\
\hline
GC01 & J & 1.25 & $4 \times 50$ & 0.052 & 0.23 & June 28 \\
 & H & 1.63 & $20 \times 10$ & 0.052 & 0.26 & June 28 \\
 & K' & 2.12 & $14 \times 35$ & 0.052 & 0.23 & June 28 \\
 & H$_2$O & 3.05 & $90 \times 2$ & 0.02 & 0.10 & July 2 \\
 & PAH & 3.30 & $100 \times 2$ & 0.02 & 0.10 & July 2 \\
 & H3+ & 3.41 & $100 \times 2$ & 0.02 & 0.10 & July 2 \\
GC02 & J & 1.25 & $4 \times 120$ & 0.052 & 0.21 & June 24 \\
 & H & 1.63 & $5 \times 60$ & 0.052 & 0.21 & June 24 \\
 & K' & 2.12 & $6 \times 60$ & 0.052 & 0.21 & June 24 \\
 & H$_2$O & 3.05 & $100 \times 2$ & 0.02 & 0.10 & July 2 \\
\hline\end{tabular}
\end{center}
\caption{Details of Observations}
\end{table*}

	As GC01 and GC02 are viewed at low Galactic latitudes, there are a number of 
potential NGSs, and this enabled the selection of a guide star 
asterism that girded the clusters, thereby increasing the chances of achieving good 
image quality. The $J, H,$ and $K'$ images were recorded with 0.052 arcsec/pixel 
sampling, with a science pick-off positioned near each cluster center. 
The GC01 field is centered at 18:48:49.9 (RA) and 
$-01:29:47.9$ (Dec), while the GC02 field is centered at 
18:18:29.7 (RA) and $-16:58:34.3$ (Dec). The co-ordinates are E2000. 
The GC01 field is within an arcsec of the cluster center, 
while the GC02 field is $\sim 5$ arcsec north of that cluster center. 

	The second pick-off sampled a background field near the edge of the RAVEN 
science field. While the background field contains some cluster 
stars, the density of objects in both GC01 and GC02 is $\sim 
50 \times$ lower than in the field sampled by the other pick-off (see Appendix). 
Star counts suggest that field stars are the dominent component at distances $> 60$ 
arcsec from the centers of both clusters (Section 3).

	One set of exposures was recorded with the cluster center positioned in or 
near pick-off \# 1, and another was recorded with the cluster centered in or near 
pick-off \# 2. While the intent was to observe the same field in each case, 
positioning uncertainties meant that there was not 100\% overlap. The stellar densities 
in the background were found to be negligible when compared with the cluster 
centers, and so the background fields are not discussed further. 

	The $J, H,$ and $K'$ data were recorded 
without on-sky dithering. While on-sky dithering can be 
employed to suppress bad pixels, pick-off \# 2 is projected onto a part of the IRCS 
detector that contains a dense collection of bad pixels, and these bad pixels 
occupy a large fraction of the science field at this pixel scale. A very large 
dither throw is required to suppress these bad pixels, and the resulting 
field coverage would then be greatly diminished. In the end, 
the pick-off \#1 images were the primary source of photometric measurements. Stars 
in portions of the pick-off \# 2 field that (1) do not overlap with the area 
sampled by pick-off \# 1, and (2) are not affected by bad pixels were 
retained for the photometric analysis of GC01. 

	The H$_2$O, PAH, and H3$+$ filters are designed to 
sample specific molecular transitions, and so have narrow bandpasses 
($0.152\mu$m for H$_2$O, $0.05\mu$m for PAH, and $0.022\mu$m for H3$+$). 
We do not use these filters to sample the intended transitions, 
but instead use them as a means to image fields near the cluster centers at wavelengths 
where better AO correction can be achieved than in the NIR, and also 
to extend the spectral-energy distribution (SED) of bright stars to wavelengths 
$> 2.5\mu$m, thereby providing greater leverage for estimating 
reddening. The narrow bandpasses of these filters help prevent saturating the 
detector given the high background levels that are inherent to these wavelengths.

	The narrow-band data were recorded with 0.02 arcsec pixel$^{-1}$ 
sampling to prevent saturating the detector with the high background levels 
that are intrinsic to ground-based observations longward of $2.5\mu$m; this pixel 
scale also allowed for better sampling of the point spread function (PSF). 
The H$_2$O, PAH, and H3$+$ images were recorded with on-sky dithering, 
as the collection of bad pixels described earlier subtends a smaller angular size on 
the pick-off \# 2 science field than is the case with 0.052 arcsec pixel sampling. 
The dither pattern consisted of five pointings that tracked the four corners 
and the center of a square-shaped asterism. An initial inspection 
of the H$_2$O observations of GC02 indicated that only a handful of 
stars could be detected with moderate exposure times in the 
longer wavelength filters, and so GC02 was observed only through the H$_2$O filter. 

	RAVEN operates at ambient temperature and contains optics that are 
optimized for wavelengths shortward of $2.5\mu$m. Thus, when observing at wavelengths 
$> 2.5\mu$m there is a high thermal background when compared with 
systems that are cryogenically cooled and have optical designs that are 
tailored to the MIR. The higher background levels hinder 
the ability to go deep at these wavelengths, although 
this can be offset in part by improvements in the delivered image quality, 
which can be significant at long wavelengths.

	Images of GC01 and GC02 in the H$_2$O filter are shown in the right hand 
column of Figure 1. The FWHM of sources in the H$_2$O images is $\sim 0.11$ arcsec, 
which is only 0.03 arcsec larger than the telescope diffraction limit at this 
wavelength. Stars are seen throughout the GC01 H$_2$O image. 
The FWHM measurements of these objects are spatially stable, with a 
dispersion of $\pm 0.01$ arcsec, which is the estimated random uncertainty 
in each FWHM measurement. The image quality of stars in GC01 when observed 
through the other narrow-band filters is also near the diffraction limit.

	$K'$ images of both clusters are shown in ther left hand column of Figure 
1. The FWHM measurements of sources in both clusters are $\sim 0.2$ arcsec in $K'$. 
Despite the larger FWHM, many more stars can be seen in 
the $K'$ images due to the lower background, coupled with the higher 
optical throughput of the system at wavelengths $> 2.5\mu$m.

\begin{figure}
\figurenum{1}
\epsscale{0.75}
\plotone{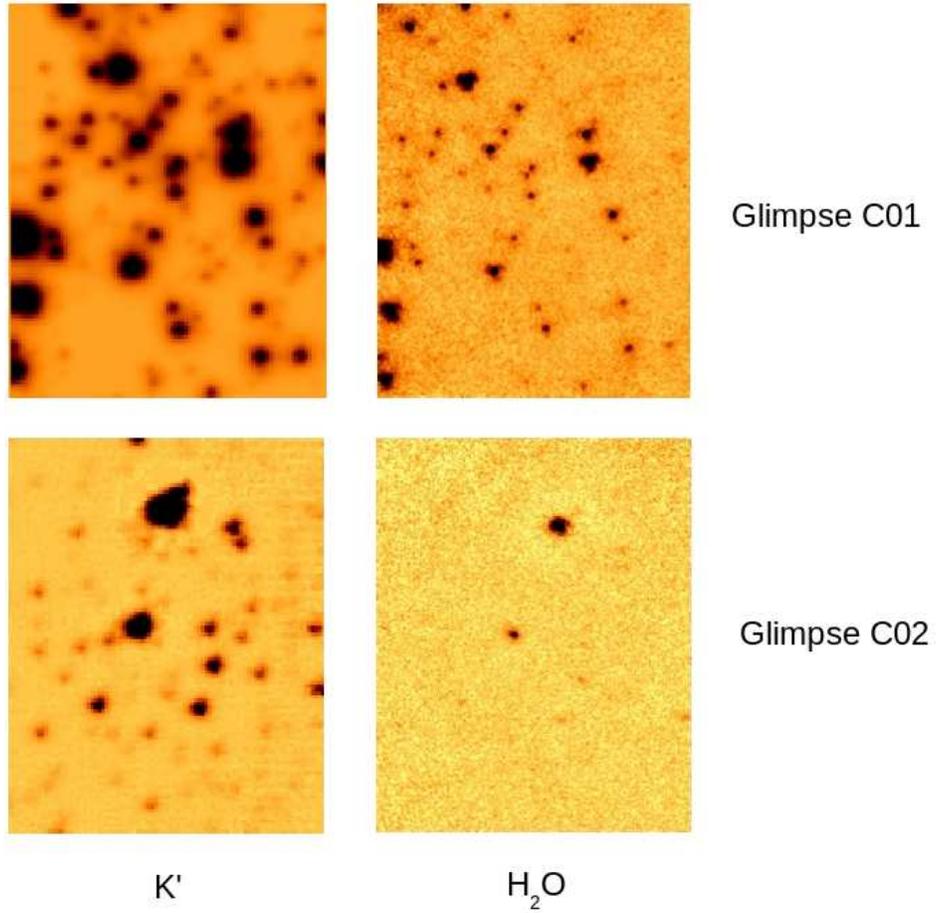}
\caption{Sections of the final processed $K'$ (left hand column) and H$_2$O 
($\lambda_{cen} = 3.05\mu$m; right hand column) images of GC01 
(upper row) and GC02 (bottom row). Each panel samples a $4.6 \times 6.1$ arcsec 
area. The faint halo around each source in the H$_2$O images is the Airy pattern.}
\end{figure}

\subsection{Processing of the Images}

	The data were reduced using a standard processing 
sequence for NIR imaging. The main steps that are applied 
prior to image combination are (1) flat-fielding, (2) the subtraction 
of a mean sky level from each image, and (3) the removal of signatures from thermally 
emitting objects along the optical path. The first step removes multiplicative 
signatures, while the second and third steps remove features that are additive.

	Pixel-to-pixel sensitivity variations and non-uniform optical throughput 
introduce multiplicative signatures into the data. 
Flat-field calibration frames were constructed from images of 
light from the telescope calibration unit that was directed into the IRCS. One 
set of exposures was recorded with the calibration light source turned on, 
and a second set was recorded with the light source turned off. 
The subtraction of the mean `off' exposure from the mean 
`on' exposure removes the additive signatures of thermal emission and dark current 
from the flat-field pattern, leaving only multiplicative effects. The calibration 
frame was normalized to unity, and the science images were divided by the result. 

	Variations in the background light level occur over short periods of time at 
these wavelengths, and these were removed before combining frames. These variations 
are due to a number of factors, including moon phase, the location of the moon 
on the sky with respect to the target, changes in 
the levels of atmospheric emission, and differential flexure between 
the telescope and instrument, which changes 
pupil mask alignment, thereby affecting the thermal background level. Pseudo-sky 
levels were found by taking the mode of the pixel intensity distribution in 
each image, and the results were subtracted from the images. The background level 
found in this way contains contributions from the night sky, the telescope and 
instrument optics, and unresolved stars in each cluster. The latter 
component means that the removal of the pseudo-sky level will negate the ability to 
measure surface brightness. However, this does not affect the ability to 
perform photometry on individual stars, as only local sky values are important.

	Images that sample wavelengths longward of $\sim 1.5\mu$m are susceptible to 
thermal emission from warm sources along the optical path, such as dust on the 
cryostat window. As these sources are out of focus, they produce diffuse 
pupil-like images. Thermal background calibration frames were 
constructed by median-combining flat-fielded images of a sky field, while 
a background calibration frame for the narrow-band observations  
was constructed for each filter using the dithered observations of GC01.
These were subtracted from the sky-subtracted flat-fielded data. 

	The processed images were aligned using a 
bright star as a reference point, and then rotated to a common 
position angle. Image registration was checked by examining the locations of stars 
throughout the field, and the alignment of the shifted/rotated images was found to 
hold at the sub-pixel level. The results from each filter$+$exposure time pair were 
combined by taking the median intensity level at each pixel location. While 
averaging the signal might produce a higher signal-to-noise ratio, 
the median is a robust means of rejecting anomalous signal that 
might arise from cosmetic defects or cosmic rays when only a modest number of 
exposures are available. 

\section{NIR PHOTOMETRIC MEASUREMENTS}

	Photometric measurements were made with the point spread 
function (PSF)-fitting routine ALLSTAR (Stetson \& Harris 1988). The PSFs, source 
catalogues, and initial brightnesses used in ALLSTAR were obtained by running tasks in 
DAOPHOT (Stetson 1987). Even though isolated stars were selected for the PSF, there 
is still contamination from faint neighbors in the crowded 
cluster fields. Contaminating stars were subtracted out by applying 
progressively cleaned versions of the PSF. The photometric 
measurements were made with a single PSF for each field$+$filter combination, 
as experiments indicated that the application of a spatially variable PSF did not 
yield tighter CMDs.

	Photometric standard stars were observed in all three NIR filters on 
four different nights (the three nights when GC01 and GC02 were observed, 
plus one additional night). The instrumental $K'$ measurements were transformed 
into $K$ magnitudes \footnote[4]{When referring to the filter we use $K'$, 
whereas when referring to photometric magnitudes obtained from the images 
we use $K$}. The zeropoints measured from the standard stars are listed 
in Table 2. The NIR zeropoints are 0.3 -- 0.4 
magnitudes brighter than the fiducial values listed on the IRCS web pages 
\footnote[5]{http://www.naoj.org/Observing/Instruments/IRCS/camera/filters.html}. 
This suggests that RAVEN has a lower thoughput than other AO systems used at Subaru, 
due in part to the transfer optics required to feed the light from the 
science pick-offs to the IRCS.

\begin{table*}
\begin{center}
\begin{tabular}{lc}
\hline\hline
Filter & Zeropoint \\
\hline
J & $25.69 \pm 0.09$ \\
H & $25.92 \pm 0.09$ \\
K & $25.22 \pm 0.09$ \\
H$_2$O & $22.10 \pm 0.1$ \tablenotemark{a} \\
PAH & $20.97 \pm 0.1$ \tablenotemark{a} \\
H3$+$ & $20.34 \pm 0.1$ \tablenotemark{a} \\
\hline
\end{tabular}
\end{center}
\tablenotetext{a}{Estimated uncertainty}
\caption{Photometric Zeropoints}
\end{table*}

	Sample completeness was estimated by running artificial star experiments.
Artificial stars were assigned colors that fall along the sequences in the 
cluster CMDs, and an artificial star was only considered to be recovered if it was 
detected in at least two filters with a maximum matching radius of 
one half the FWHM. The dispersions in the recovered magnitudes and 
completeness fractions were computed after applying an iterative $2.5 \sigma$ 
rejection filter to the mean difference between input and measured 
brightnesses in 0.5 magnitude intervals. With the exception of the $J$ observations 
of GC02, the magnitude at which incompleteness sets in is defined by 
crowding in the NIR data, rather than photon statistics. The photometric 
faint limits are thus much brighter than what would otherwise be expected from 
images recorded with an 8 meter telescope.

\subsection{GC01}

	The $(K, J-K)$ and $(K, H-K)$ CMDs of GC01 are shown in Figure 2. 
The stars plotted in these CMDs were matched in filter pairs (i.e. $J$ and $K$ or 
$H$ and $K$, depending on the CMD), rather than requiring a match in all 
three filters. A maximum matching radius of one-half of the FWHM of the 
wider of the two PSFs was adopted -- sources in one filter that did not have a 
match within this radius are not included in the CMDs. 
The 50\% completeness levels determined from the artificial star experiments are 
indicated. 

\begin{figure}
\figurenum{2}
\epsscale{1.00}
\plotone{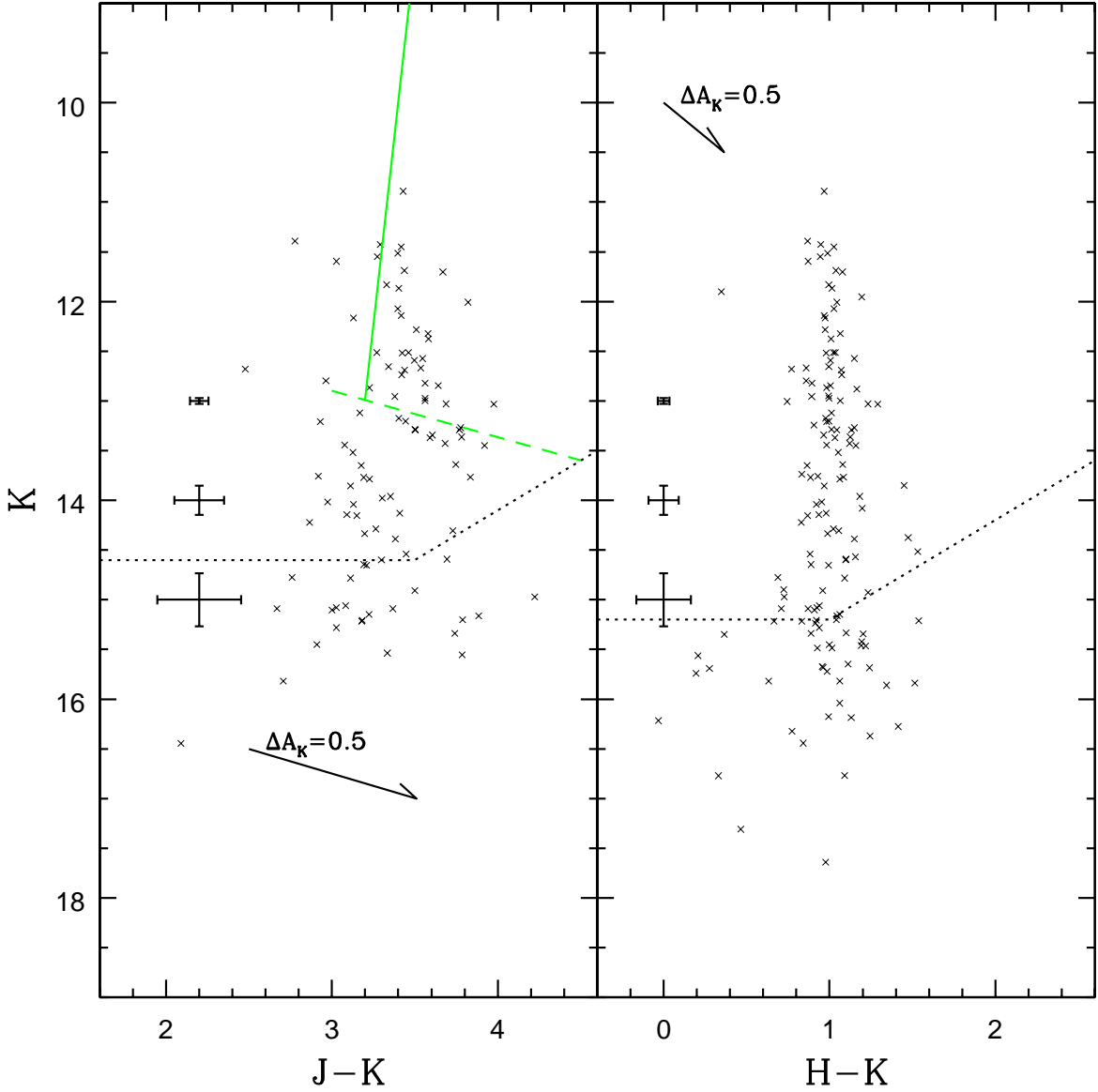}
\caption{$(K, J-K)$ and $(K, H-K)$ CMDs of GC01. The error bars show the 
$\pm 1 \sigma$ dispersion in the photometry calculated from the artificial 
star experiments. The dotted black lines are the 50\% completeness 
limits, also from the artificial star experiments. The 
observations are $\sim 100\%$ complete 0.5 magnitudes above the 50\% 
completeness limit. The solid green line is the fiducial 
giant branch sequence from the middle panel of Figure 4 of Ivanov et al. (2005), 
while the dashed green line is a hand-drawn fit to the RC sequence in that same 
figure. Reddening vectors with lengths that correspond to $\Delta$A$_K = 0.5$ 
and that follow the Nishiyama et al. (2009) reddening law are also shown.}
\end{figure} 

	The error bars in Figure 2 show the $\pm 1\sigma$ uncertainties 
calculated from the artificial star experiments. 
The scatter near the faint end of the CMDs more-or-less matches the error bars. 
However, the scatter near $K \sim 13$ in the $(K, J-K)$ CMD of GC01 
exceeds that expected from random photometric uncertainties, 
and we attribute this to differential reddening. 
A reddening vector, with an amplitude corresponding to $\Delta$A$_K = 0.5$ 
magnitudes, is shown in each panel of Figure 2, and it can be seen that $\Delta$A$_K$ 
of a few tenths of a magnitude can explain the scatter in the $(K, J-K)$ CMD near 
$K = 13$. This scatter prevents us from measuring the slope of the giant 
branch, which might otherwise be used to estimate metallicity. 

	The fiducial giant branch sequence from the middle panel of Figure 4 of 
Ivanov et al. (2005) is shown as a solid green line in Figure 2. The Ivanov 
et al. (2005) photometry is in the $Ks$ filter system, and so their measurements may 
differ from those in $K$ by up to a few hundredths of a magnitude (e.g. Table 2 
of Persson et al. 1998). The Ivanov et al. (2005) fiducial skirts the blue edge of 
the IRCS$+$RAVEN CMD. Assuming that the location of the Ivanov et al. (2005) fiducial 
indicates a lower mean extinction in the outer regions of GC01 then -- based on the 
Nishiyama et al. (2009) reddening law -- the typical A$_K$ towards the center 
of GC01 is $\sim 0.15$ magnitudes higher than at larger radii. This higher 
reddening is perhaps not surprising given the broad range in A$_K$ that is found 
throughout GC01, coupled with the warm dust lane that cuts through the cluster center 
(Kobulnicky et al. 2005). 

	Ivanov et al. (2005) identify a concentration of 
stars near $K \sim 13$ in their CMD of objects within 20 arcsec of the cluster center 
that they suggest is the red clump (RC). The RC in the left hand panel of their 
Figure 4 forms a tilted sequence -- likely due to differential reddening -- and 
the locus of this sequence is shown in Figure 2. There is not an obvious corresponding 
sequence in the IRCS$+$RAVEN $(K, J-K)$ CMD. Nevertheless, 
a concentration of stars due to the RC appears in the CMD after 
correcting for differential reddening (see below).

	The NIR SED of stars near the center of 
GC01 is examined in Figure 3, where the $(J-H, H-K)$ two-color diagram (TCD) 
is shown. The dotted line is the locus of points in Figure 5 
of Ivanov et al. (2005), and there is good agreement with the RAVEN 
measurements. Fiducial sequences for red giants and Iab supergiants 
from Bessell \& Brett (1988) are also shown in Figure 3, as is a reddening vector 
that tracks the Nishiyama et al. (2009) reddening law. 
Differences between various reddening laws become 
significant for highly obscured objects like GC01. Extrapolating the GC01 observations 
along the Nishiyama et al. (2009) relation comes closer to 
intersecting the area of the TCD that contains the unreddened colors of giants than 
extrapolating along vectors defined by the Rieke \& Lebofsky (1985) and the 
R$_V = 3.1$ Cardelli et al. (1989) reddening laws. The Nishiyama et al. (2009) 
reddening law is thus adopted for the remainder of the paper.

\begin{figure}
\figurenum{3}
\epsscale{1.00}
\plotone{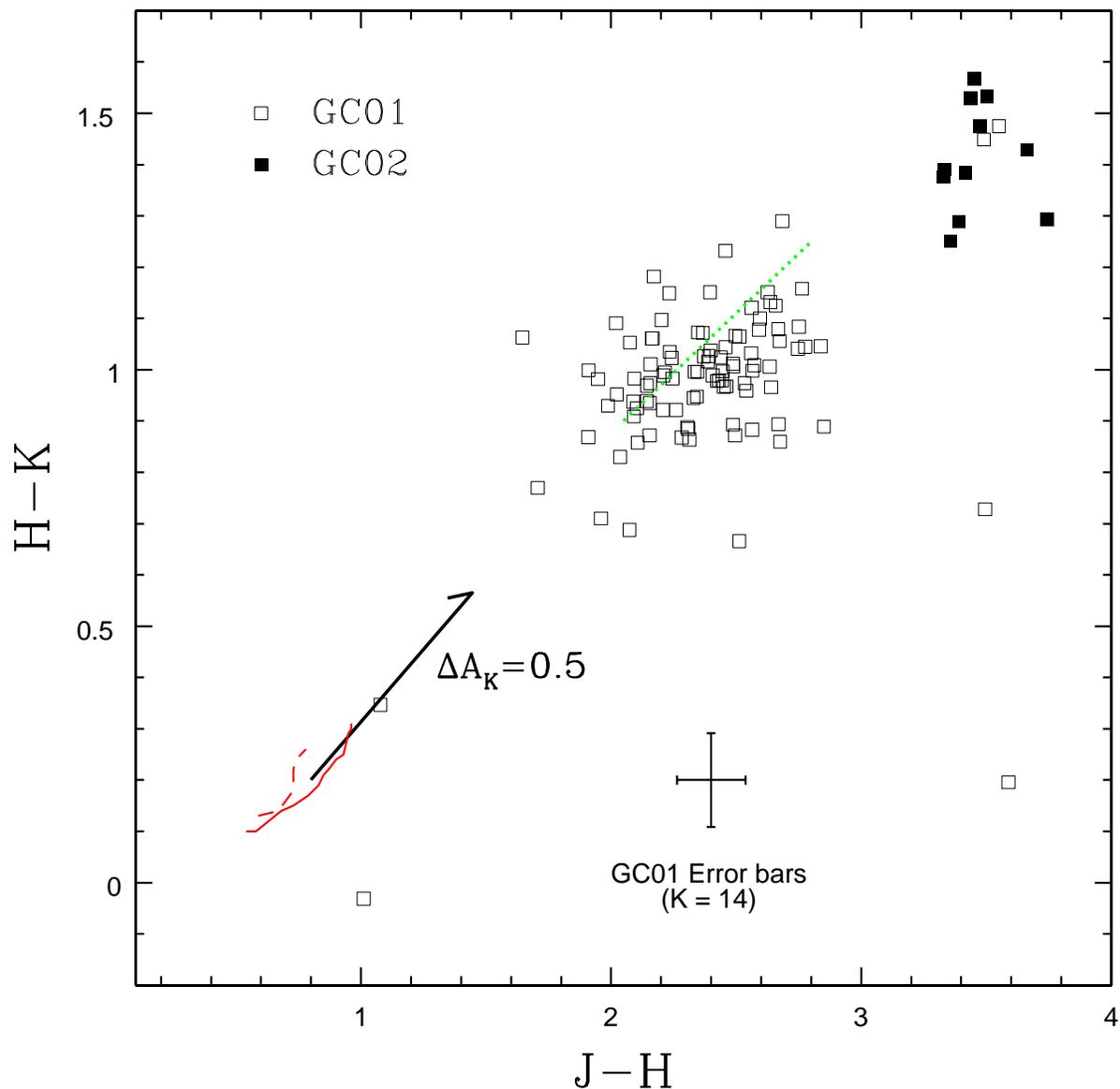}
\caption{$(J-H, H-K)$ two color diagram of stars near the centers of GC01 
(open squares) and GC02 (filled squares). The error bars show the 
$\pm 1 \sigma$ uncertainties estimated from artificial star 
experiments for objects in GC01 with K = 14. 
The solid and dashed red lines are fiducial sequences for Galactic giants and Iab 
supergiants from Bessell \& Brett (1988). A reddening vector that follows 
the Nishiyama et al. (2009) reddening law and has a length that corresponds 
to $\Delta$A$_K = 0.5$ magnitudes is shown. The trend defined by this vector 
links the fiducial and GC01 observations. The dotted green line is the locus of 
points plotted in Figure 5 of Ivanov et al. (2005).} 
\end{figure}

	Under ideal circumstances, reddenings for individual stars could be estimated 
by projecting each point on the TCD back to the unreddened giant sequence. 
However, there are large (when compared with color differences between 
stars with different spectral-types) random uncertainties in the photometric 
measurements. Photometric variability contributes further 
to the smearing, although Adelman (2001) finds that variability is likely 
not a factor among RC stars, at least over time scales of a few years. 
As scatter in the observations impedes efforts to identify an intrinsic color 
for individual stars, reddenings are estimated here by assuming that 
the stars in the GC01 CMDs have a common intrinsic color. 
This assumption is reasonable as our CMDs sample giants with M$_K$ between 0 and -3, 
and solar metallicity isochrones generated from the Marigo et al. (2008) models 
indicate that a narrow range of spectral-types (K2III to K5III) 
is expected in this M$_K$ interval in old systems. 

	If the stars in the CMDs are assumed to have the intrinsic $J-H$ and $H-K$ 
colors of a K3 III star then the mean reddening towards the center of GC01 
is A$_K = 1.24 \pm 0.01$ based on the Nishiyama et al. (2009) reddening 
law. The uncertainty is the standard error of the mean. 
The reddening depends on the adopted intrinsic colors, and 
if it is assumed that the intrinsic NIR colors of each star match those 
of -- say -- an M1 III giant, which falls near the middle of the giant sequence in 
Figure 3, then the mean extinction would be A$_K = 1.13 \pm 0.01$. 

	The $1 \sigma$ dispersion in A$_K$ computed from the TCD 
is $\pm 0.13$ magnitudes. Given the scatter due to photometric errors and 
the assumption that all stars have the same intrinsic color then 
this is likely an upper limit to the smearing caused by differential reddening. 
Sources with A$_K$ between 1 and 1.5 are well-mixed throughout 
the RAVEN$+$IRCS field, suggesting that significant variations in line-of-sight 
reddening towards stars in GC01 occur over sub-arcsec angular scales. 
An arcsec corresponds to a spatial scale of $\sim 0.025$ parsecs 
at the distance of GC01. Given that the obscuring material is either at 
the distance of GC01 or is in the foreground then the ISM towards 
the center of GC01 contains structure over spatial scales of no more 
than $\sim 10^4$ AU.

	The CMDs constructed from the unreddened photometric measurements are shown 
in the top panels of Figure 4. The vertical sequences in these CMDs are the 
direct result of adopting a single intrinsic color when computing reddenings. 
The number of sources in the CMDs in Figures 2 and 4 are not 
the same, as only stars that were detected in all 
three NIR filters have been de-reddened. Hence, the CMDs in Figure 4 contain 
fewer stars than those in Figure 2.

	While the assumption of a common NIR SED for all stars suppresses 
color-related information in the reddening-corrected photometry, the 
reddening-corrected CMDs still contain useful information. There is a local peak in the 
number of stars near K$_0 \sim 11.9 - 12.0$ in Figure 4, which we identify as the 
RC. These stars have $K = 13.1 - 13.2$ if $<A_K> = 1.24$, which 
agrees with the RC magnitude found by Ivanov et al. (2005) at larger radii. There is 
also a drop in the number of stars in the $\sim 0.5$ magnitude interval 
fainter than the RC. The artificial star experiments indicate that the data 
are complete to K$_0 \sim 13.4$, and so the drop in number counts immediately 
below the RC in the CMD is not due to sample incompleteness.

	The change in number counts near K$_0 = 12$ is examined in the 
lower panel of Figure 4, where cumulative number counts in 0.2 magnitude 
intervals are shown. The green dashed line is a least squares fit to the 
cumulative counts with $K_0 \leq 12$. The rate of growth in number counts 
changes significantly near K$_0 = 12$. Models of stellar evolution predict such a 
change to occur at magnitudes below the RC (see below).

\begin{figure}
\figurenum{4}
\epsscale{1.00}
\plotone{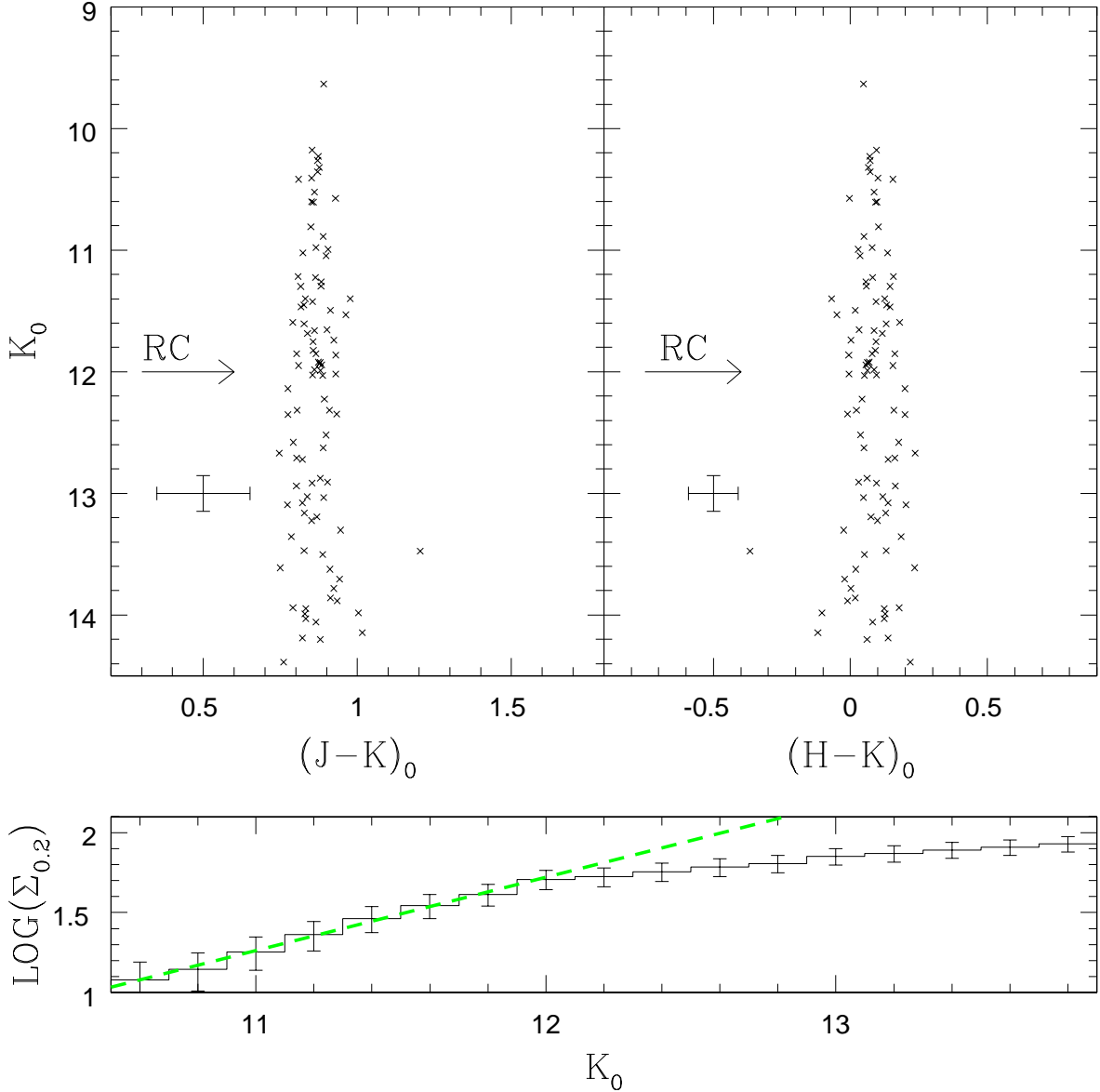}
\caption{(Top panels) Reddening-corrected $(K, J-K)$ and $(K, H-K)$ CMDs of GC01. The 
subscript `0' denotes magnitudes and colors that have been 
de-reddened according to their location on the $(J-H, H-K)$ diagram and assuming 
intrinsic colors matching those of a K3 giant. The $\pm 1\sigma$ error bars 
for K$_0$=13 from the artificial star experiments are shown. The vertical sequences 
result from the assumption that all stars have the same intrinsic colors. 
The concentration of objects near $K = 12$ is due to the red clump. 
(Bottom panel) Cumulative number counts from the CMDs in the top panels. 
$\Sigma_{0.2}$ is the sum of the number of sources in 0.2 magnitude bins up to a given 
magnitude. The green dashed line is a least squares fit to the entries 
with K$_0 \leq 12$. The rate of growth of the cumulative 
number counts changes near K$_0 = 12$, which is the brightness of the RC.}
\end{figure}

	Van Helshoecht \& Groenewegen (2007) examine the brightness 
of the RC in clusters that span a range of metallicities and ages. They conclude 
that M$_K$(RC)$= -1.57 \pm 0.05$ in systems that have metallicities within a few tenths 
of a dex of solar and ages between 0.3 and 8 Gyr. Assuming that GC01 falls 
within this age range and has a near-solar metallicity, then it has a distance 
modulus $= 13.6$, corresponding to a distance of 5.2 kpc.

	The $K$ LF of GC01, constructed from the de-reddened 
$(K, H-K)$ CMD, is shown in Figure 5. The range of magnitudes has been 
restricted to those where artificial star experiments predict that the data are 
complete. A 0.5 magnitude bin width was adopted as it is the smallest that would 
produce meaningful numbers of stars per bin in this magnitude range. The 
discussion that follows will not change significantly if a different starting 
point for binning is adopted. The models that are compared with the GC01 LF (see below) 
were constructed using the same binning parameters as the observations
in an effort to further mitigate against binning errors. 

\begin{figure}
\figurenum{5}
\epsscale{0.90}
\plotone{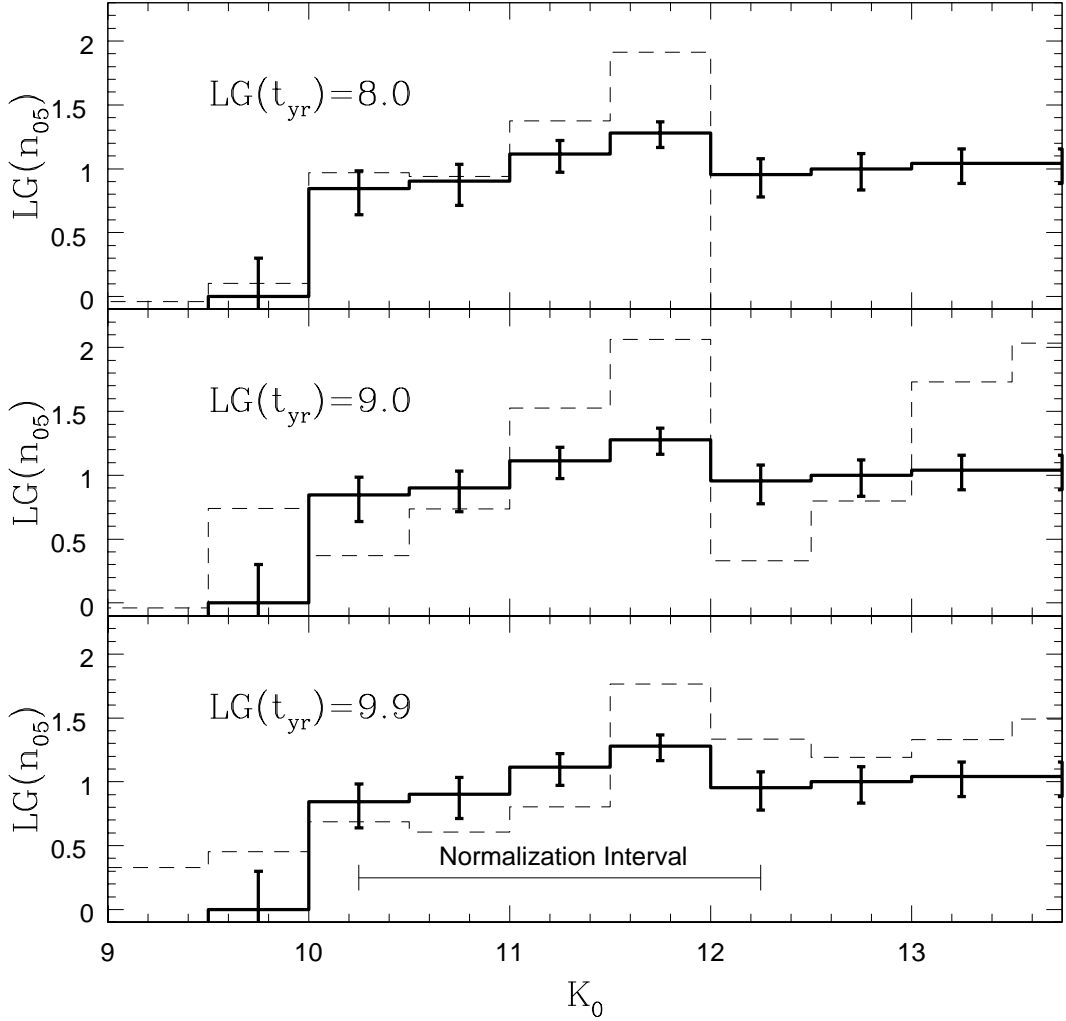}
\caption{$K_0$ LF of GC01, with number counts from the de-reddened CMDs shown in 
Figure 4. $K_0$ is the de-reddened $K$ magnitude, 
while n$_{0.5}$ is the number of stars per 0.5 magnitude 
interval in $K$. Artificial star experiments indicate that the data are 
complete to $K_0 = 13.5$, and the completeness in the faintest bin is $\sim 50\%$. 
The error bars show $\pm 1 \sigma$ uncertainties calculated from counting statistics. 
Model LFs constructed from solar metallicity Marigo et al. (2008) isochrones with 
ages log(t$_{yr}$) = 8.0, 9.0, 9.9 are shown as dashed lines. 
The models have been normalized to match the number of sources with K$_0$ between 
10 and 12. This magnitude interval contains many of the detected stars, 
and samples magnitudes where the number counts are complete. The youngest models 
do not match the overall trends in the LF: the log(t$_{yr}$) = 8 model underestimates 
the number of sources with K$_0 > 12$ by more than an order 
of magnitude, whereas the log(t$_{yr}$) = 9.0 model overestimates 
the amplitude of the RC. The log(t$_{yr}$) = 9.9 model matches the 
amplitude of the discontinuity near $K_0 = 12$ and yields the best overall 
match with the observations, although the number of stars with K$_0$ between 10 
and 11.5 is underestimated by a factor of two.}
\end{figure} 

	The shape of the LF in Figure 5 provides clues to the age of GC01, 
although features in LFs can be affected by 
stellar variability and uncertainties in the line-of-sight 
extinction, both of which cause smearing along the magnitude axis. 
Davidge (2000) discusses the $K$ LF of the metal-rich globular 
cluster NGC 6528, and those data provide an empirical point of comparison 
for GC01. The LF of NGC 6528 shown in Figure 3 of 
Davidge (2000) climbs towards fainter magnitudes with the RC 
forming a pronounced peak. A smaller peak due to the RGB-bump (Iben 1968) is seen 
$\sim 1$ magnitude fainter than the RC. Finally, there is a marked 
jump in the NGC 6528 LF $\sim 2$ magnitudes fainter than the RC that is due to 
the onset of the sub-giant branch (SGB). 

	There are similarities and differences between the GC01 and NGC 
6528 LFs. The amplitude of the RC with respect to fainter 
stars in GC01 is comparable to that in NGC 6528. 
However, when considered over a wide range of magnitudes 
the LF of GC01 is more-or-less flat, and the 
ratio of stars that are brighter than the RC to those that are fainter 
is higher in GC01 than in NGC 6528. Unfortunately, the GC01 data 
do not go faint enough to sample the magnitude where the onset of the SGB 
occurs in NGC 6528.

	Model LFs of simple stellar systems 
(SSPs) constructed from the Marigo et al. (2008) isochrones are compared with the 
GC01 LF in Figure 5. The distance modulus applied to each model was set to match the 
brightness of the RC predicted by that model. The models have been scaled 
along the vertical axis to match the observations between K$_0 = 10$ and 12. 
This magnitude interval contains the majority of detected stars and is 
where the sample is statistically complete.

	The amplitude of the RC in the LF contains information about age. The 
models in Figure 5 demonstrate that the amplitude of the drop to the faintward side of 
the RC is age-sensitive. Similar behaviour can be seen in the compilation of open 
cluster CMDs examined by Van Helshoecht \& Groenewegen (2007) that were used to 
establish their M$_K$ calibration of the RC. The 9 clusters in their Figure 8 that show few if any stars faintward of the RC (IC4651, NGC 2090, NGC 2380, NGC 2477, NGC 2527, 
NGC 3680, NGC 3960, NGC 5822, and NGC 7789) have a mean age log(t$_{yr}$) $= 8.87 \pm 
0.07$, where the uncertainty is the formal error of the mean and the ages are taken 
from Table 3 of Van Helshoecht \& Groenewegen (2007). In contrast, the mean age of the 
clusters that have a well-defined sequence faintward of the RC (Be 39, Mell 66, NGC 
188, NGC 1817, NGC 2243, NGC 2506, NGC 2582, NGC 6633, NGC 6791, NGC 6819) is 
log(t$_{yr}$) $= 9.25 \pm 0.12$. The difference in mean age between these 
two groups is significant at the $2.7\sigma$ level.

	The LFs of the log(t$_{yr}$)=8 and log(t$_{yr}$)=9 populations greatly 
over-estimate the amplitude of the drop in the LF faintward of $K_0 = 12$, although 
the former model matches the LF shape and number of stars when K$_0 < 11$. In both 
cases the difference between the K$_0 = 12$ and 12.5 bins is exceeded by the 
models at more than the $10\sigma$ level. The log(t$_{yr}$) = 9 
model also predicts a steep rise in the number counts 
at magnitudes brighter than the RC and a large increase in number counts at K$_0 = 13$ 
due to the onset of the main sequence. Corresponding features are not seen in 
the observations. 

	The log(t$_{yr}$)=9.9 model matches best the entire LF, although the agreement 
is far from ideal as the model does not reproduce the overall flat 
nature of the GC01 LF. In fact, the bright portions of the GC01 LF show 
similarities to the log(t$_{yr}$)=8 model, while the fainter portions show properties 
that are consistent with the log(t$_{yr}$) = 9.9 model. While none of the models 
provide an ideal match with the observations, the comparisons in Figure 5 suggest that 
GC01 might contain a sizeable population of stars with log(t$_{yr}$) $> 9$ based on 
the drop in number counts faintward of the RC.

\subsection{GC02}

	The $(K, J-K)$ and $(K, H-K)$ CMDs of GC02 are shown in Figure 6. 
The CMDs were restricted to sources imaged in pick-off \# 1, as only a single 
exposure per filter was recorded with the cluster centered in pick-off \# 2, 
thereby preventing the suppression of bad pixels. The error bars 
show the $\pm 1 \sigma$ dispersion predicted from the 
artificial star experiments.

\begin{figure}
\figurenum{6}
\epsscale{1.00}
\plotone{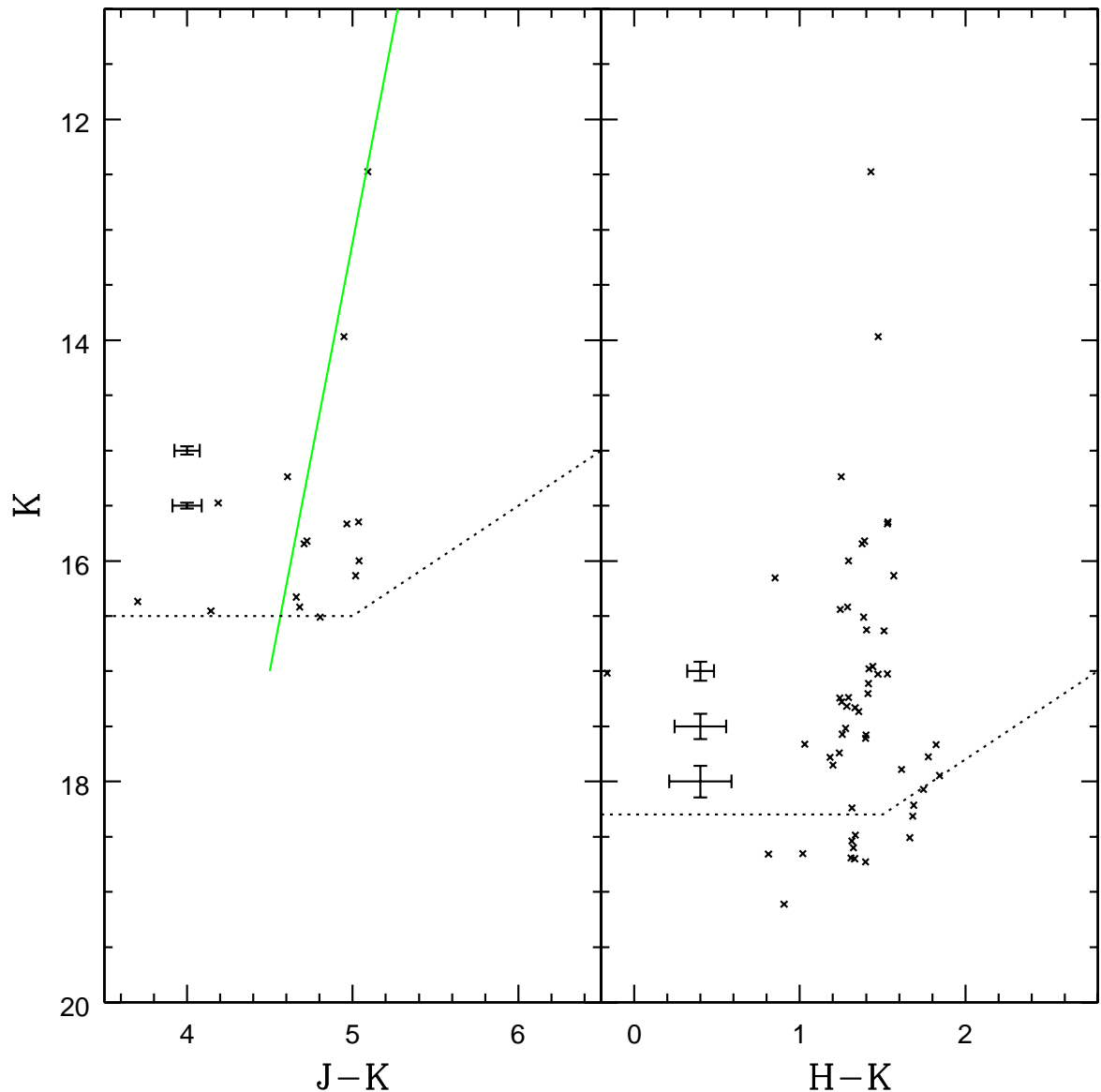}
\caption{The $(K, J-K)$ and $(K, H-K)$ CMDs of GC02. The green solid line is the 
cluster ridgeline from Figure 2 of Kurtev et al. (2008). The dotted lines are the 
50\% completeness levels estimated from the artificial star experiments, 
while the error bars show the predicted $\pm 1 \sigma$ dispersions in the 
photometric measurements, also from the artificial star experiments. The $(K, 
H-K)$ CMD of GC02 goes $\sim 2 - 3$ magnitudes fainter than the GC01 CMD in Figure 2
because of the lower stellar density at a given $K$ in GC02.}
\end{figure}

	The $(K, J-K)$ CMD contains only a modest number of objects, 
due to the high line-of-sight extinction towards GC02, which limits 
the depth of the $J$ observations. The solid line on the $(K, J-K)$ CMD is the 
fiducial sequence from Figure 2 of Kurtev et al. (2008). This 
relation passes through the points in our CMDs, suggesting that the 
reddening towards the center of GC02 is similar to that at larger radii. 
This agreement also suggests that -- unlike GC01 -- there is probably 
not substantial differential reddening near the center of GC02. 

	The $(K, H-K)$ CMD of GC02 is more richly populated than the 
$(K, J-K)$ CMD, owing to the lower line-of-sight extinction in $H$ when compared 
with $J$. If A$_K \sim 1.7$ mag (see below) then 
the total extinction in $J$ as $\sim 2.5$ magnitudes higher than in $H$, 
and this accounts for the difference in 50\% completeness levels between the two 
CMDs in Figure 6. The scatter in the $(K, H-K)$ CMD is comparable 
to that in the GC01 CMD in Figure 2, although the GC02 $(K, H-K)$ CMD goes 2 - 3 
magnitudes deeper. The difference in photometric depth is due to 
the lower density of sources at a given $K$ in GC02, with the result that crowding sets 
in at a fainter magnitude than in GC01. In the Appendix it is shown that the 
[3.6] surface brightness near the center of GC02 is $\sim 2.5$ magnitudes/arcsec$^2$ 
lower than in GC01. If it is assumed that the two clusters have the same distances 
and that their LFs have the same shape -- but are scaled according to surface 
brightness -- then this lower surface brightness can account for much of the 
difference in depths between the GC01 and GC02 observations.

	The locations of points in GC02 on the $(J-H, H-K)$ TCD is shown in Figure 3. 
As was the case with GC01, the Nishiyama et al. (2009) reddening vector 
links the fiducial and observed sequences. Applying the procedure discussed in 
Section 3.1, the mean reddening based on the TCD is $<A_K> = 
1.95 \pm 0.03$ if the stars have a K3III spectral-type. This is based on only 
a handful of measurements, and an estimate that involves more points can be obtained 
using the mean color in the $(K, H-K)$ CMD. Assuming a K3III spectral-type 
then E(H-K) = 1.26, so that A$_K \sim 1.73$, with an estimated uncertainty of 
$\pm 0.1$ magnitude. This reddening is adopted for the remainder of the paper given 
the larger number of points involved in its calculation. If the RC occurs near 
$K = 14.4 \pm 0.15$ (Kurtev et al. 2008) then the distance modulus of GC02 is $14.2 \pm 
0.2$, corresponding to a distance of $6.9 \pm 0.5$ kpc. As with GC01, the RC magnitude 
calibration from Van Helshoecht \& Groenewegen (2007) has been adopted, and 
the $\pm 0.05$ magnitude uncertainty in the M$_K$ of the RC 
has been included when calculating the uncertainty in the distance.

	The majority of stars in the CMDs are fainter than the RC, 
indicating that IRCS$+$RAVEN may have detected stars in GC02 that are 
evolving on the lower giant branch. This opens the possibility of estimating an age 
based on the presence/absence of the SGB. The $K$ LF of GC02, with number 
counts taken from the $(K, H-K)$ CMD, is shown in Figure 7. 
There is an increase in number counts between 
$K = 15$ and 17, at which point the LF levels off. There are 
no stars detected near $K = 14.4$, likely 
due to the modest density of stars in this field. 
The LF of GC02 thus differs from that of GC01 in Figure 5, which 
is flat in the 1.5 magnitude interval fainter than the RC.

	Kurtev et al. (2008) suggest that GC02 is an old metal-rich globular 
cluster. As such, it should have a LF that is similar to those of the old, 
metal-rich clusters NGC 6528 and Liller 1. However, the onset of the SGB that 
occurs $\sim 3$ magnitudes in $K$ fainter than the RC in the LFs of the 
metal-rich globular clusters NGC 6528 and Liller 1 that are shown in 
Figure 3 of Davidge (2000) is not evident in the GC02 LF. The absence of a SGB could 
indicate that the distance modulus of GC02 is in error, although in Section 4.2.2 
it is shown that the brightness of the RGB-tip in GC02 measured from 
SPITZER images is consistent with that found by Kurtev et al. (2008).

\begin{figure}
\figurenum{7}
\epsscale{1.00}
\plotone{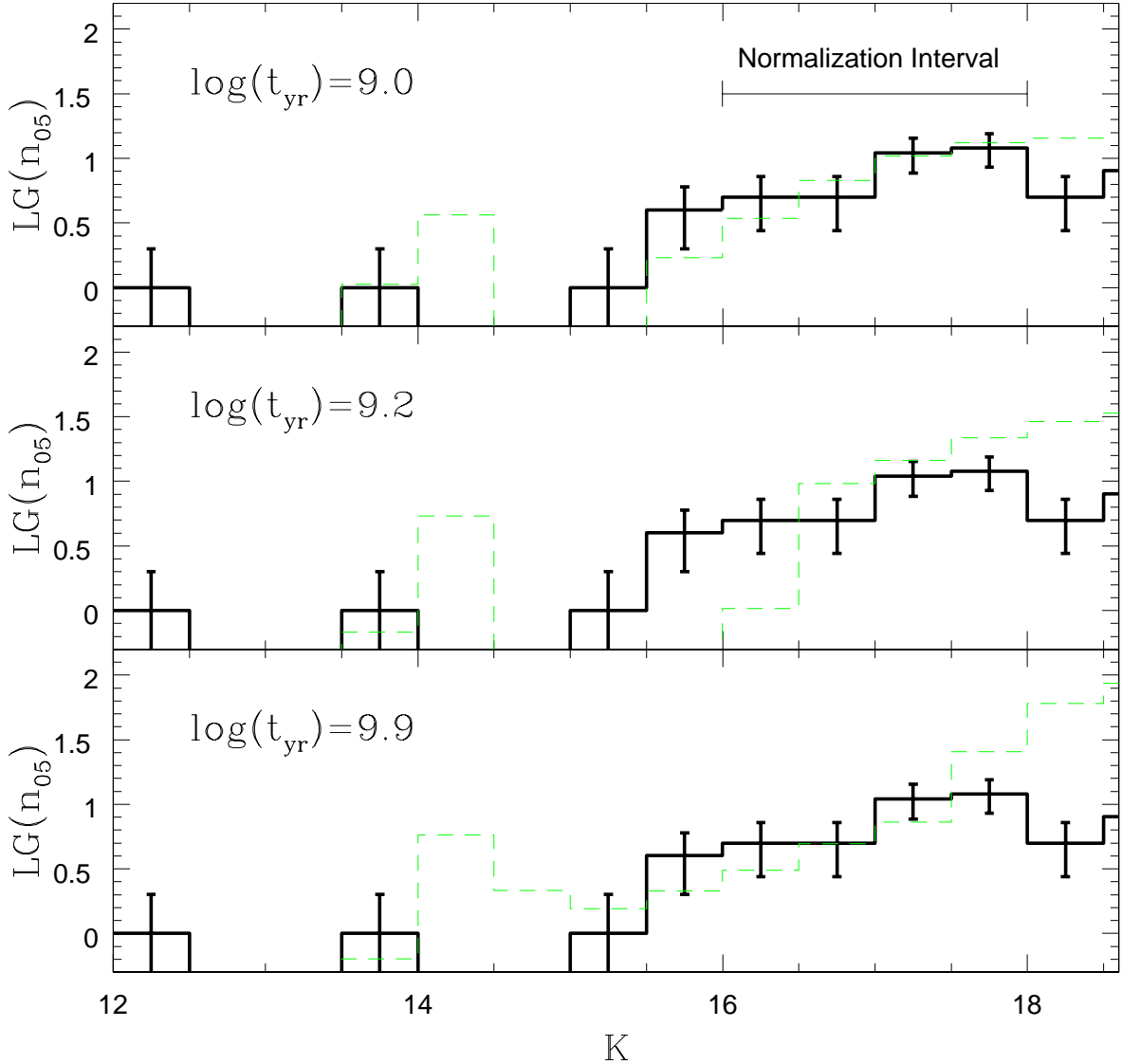}
\caption{$K$ LF of GC02, with number counts taken from the $(K, H-K)$ CMD. 
$n_{0.5}$ is the number of points per 0.5 magnitude interval in $K$. 
The errorbars show $\pm 1 \sigma$ uncertainties computed using counting statistics. 
Artificial star experiments indicate that the data are 50\% complete in the 
last bin, and are complete in the brighter bins. The 
green lines are model LFs constructed from the solar metallicity Marigo et al. 
(2008) isochrones with log(t$_{yr}$) = 9.9, 9.2, and 9.0. The models have been 
normalized to match the observations between $K = 16$ 
and 18. The log(t$_{yr}$) = 9.2 and 9.9 models assume a distance modulus of 14.2
whereas the log(t$_{yr}$) = 9.0 model has been assigned a distance modulus of 13.7 to 
force the model brightness of the RC to occur at $K = 14.4$.}
\end{figure}

	Another possibility is that GC02 may have an age that is 
very different from that of NGC 6528 and Liller 1. Model LFs 
constructed from solar-metallicity isochrones from Marigo et al. (2008) are 
compared with the observations in Figure 7. The log(t$_{yr}$) = 9.2 and 9.9 
models are shown for a distance modulus of 13.7, and these 
predict that the RC occurs in the $K = 14$ to 14.5 
interval, as observed. However, the RC in the log(t$_{yr}$) = 9.0 model 
occurs $\sim 0.5$ magnitudes fainter than found by Kurtev et al. (2008) 
if the distance modulus is 13.7. A distance modulus of 13.2 was thus assumed 
for this model to force agreement with the observed magnitude of the RC.

	There are sizeable error bars at all magnitudes in Figure 7, 
and the modest number of stars means that the amplitude of 
the RC with respect to stars in adjacent magnitude bins, 
which was used to explore the age of GC01, can not be used to constrain the age 
of GC02. Neither the log(t$_{yr}$) = 9.2 or 9.9 models match the overall shape of 
the LF. While the log(t$_{yr}$) = 9.9 model agrees with the number counts between 
$K = 15$ and 17, the model counts climb when $K > 17.5$ 
due to the onset of the SGB, and this is not seen in the 
observations. A similar disagreement is also seen near the faint end of the 
log(t$_{yr}$) = 9.2 model.

\section{SPITZER AND MIR NARROW-BAND PHOTOMETRY}

	Observations at wavelengths $> 2.5\mu$m provide 
information about the properties of late-type cluster members. 
The line-of-sight extinction at these wavelengths is lower than at shorter wavelengths, 
and -- when compared with the NIR and visible regions -- there is also 
improved contrast between the reddest stars and the (bluer) main body of the cluster. 
This raises the possibility that bright red stars 
in the dense central cluster regions might be resolved with only minimal contamination 
from bluer, intrinsically fainter cluster members. Finally, the SEDs of 
stars at these wavelengths provide checks on the reddenings measured at shorter 
wavelengths. 

	Two datasets are used in this Section to examine the photometric properties 
of stars in GC01 and GC02 at wavelengths longward of $2.5\mu$m. 
One dataset consists of the narrow-band images 
that were recorded with RAVEN$+$IRCS and 
were described in Section 2. While having a modest 
science field, these images have angular resolutions that 
approach the diffraction limit of an 8 meter telescope, and so provide checks 
on crowding among bright red stars in datasets that have poorer angular resolutions. 
These data are also used to extend the SEDs of bright stars in both clusters 
to wavelengths longward of $2.5\mu$m, and are used to check reddening.

	The other dataset consists of [3.6] and [4.5] 
images that were recorded as part of the GLIMPSE survey. The 
SPITZER observations cover a large area on the sky, allowing a comprehensive census 
of the brightest stars in and around the clusters. 
Details of the GLIMPSE survey are discussed by Benjamin et al. (2003). 
The survey was conducted in all four IRAC bands with an exposure time per 
1.2 arcsec pixel of 2 seconds. The images used here were extracted from post-basic 
calibrated data (PBCD) mosaics that have been re-sampled to 0.6 arcsec pixels. 
$0.3 \times 0.3$ degree sections of the PBCD mosaics that are centered on 
both clusters were downloaded from the NASA/IPAC Infrared Science Archive 
\footnote[6]{http://irsa.ipac.caltech.edu/data/SPITZER/GLIMPSE/}.
The angular resolution of the [3.6] and [4.5] observations is $\sim 1.7$ arcsec FWHM 
(Fazio et al. 2004), potentially complicating efforts to resolve individual stars 
near the cluster centers. A [3.6] image of each cluster is shown in Figure 8.

\begin{figure}
\figurenum{8}
\epsscale{0.80}
\plotone{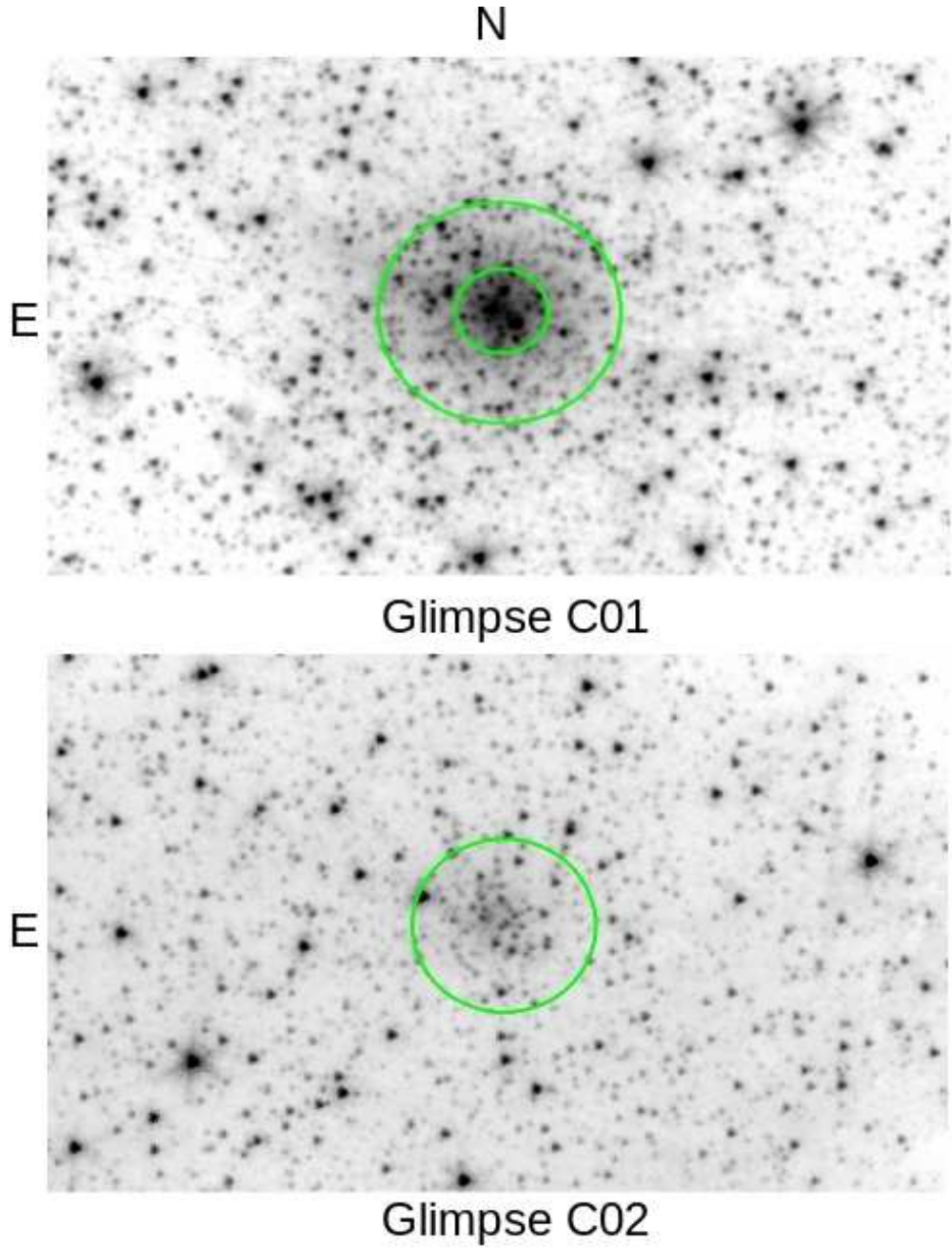}
\caption{[3.6] images of GC01 and GC02, extracted from PBCD mosaics 
constructed from GLIMPSE survey images. Each panel subtends $225 \times 
400$ arcsec, with North at the top, and East to the left. The green circles in the 
GC01 panel show the 18 -- 48 arcsec annulus that is used to extract a sample 
of cluster stars in Section 4.1.2. The green circle in the GC02 panel shows the 
area with an outer radius of 36 arcsec that is used to extract a sample of stars for 
that cluster.}
\end{figure}

	Photometric measurements of the SPITZER images of both clusters and 
of the RAVEN GC01 observations were made with 
ALLSTAR (Stetson \& Harris 1988), with PSFs constructed using the procedures 
described in Section 3. Because of the low stellar density, stellar brightnesses in 
the GC02 RAVEN H$_2$O data were measured with the PHOT routine in DAOPHOT (Stetson 
1987). The SPITZER photometry was calibrated using the 
zeropoints listed in Table 7 of Reach et al. (2005). The calibration of 
the narrow-band measurements is based on observations of Gliese 
748 (Gl748). The SED of Gl748 in the $2.0 - 3.5\mu$m interval 
is assumed to follow that of Gl273, which has the same spectral-type as Gl748 (M3.5V) 
and has flux densities tabulated by Rayner et al. (2009).
The magnitudes measured from the narrow-band observations are in the 
AB system (Oke \& Gunn 1983). 

	Gl748 is a binary system, with a component separation of 0.1 - 0.2 
arcsec, that is listed as a photometric standard by Elias et al (1982) and Leggett et 
al. (2003). However, after the RAVEN$+$IRCS data were obtained, we became aware of the 
work of Franz et al. (1998), who found that the difference in $V$ magnitude between 
the two components differed by up to 0.24 magnitudes 
over a 2.5 year period. The standard deviation of the magnitude difference 
obtained over 14 different epochs is $\pm 0.09$ magnitudes.
While this is a source of concern, the narrow-band filters sample the tail end 
of the SED of both components of Gl748, and so the SED shape -- and hence color -- 
of Gl748 at wavelengths $> 3\mu$m likely does not vary significantly with time. 

	The uncertainty in the photometric calibration of the narrow band measurements 
is estimated to be $\pm 0.1$ magnitudes, and the zeropoints are shown in Table 2. 
This calibration of the narrow-band photometry indicates that the overall throughput 
of RAVEN$+$IRCS drops considerably with wavelength when $\lambda > 2.5\mu$m. We note 
that the RAVEN optics were not designed to work at these wavelengths, and so 
the poor throughput is not a surprise. 

\subsection{Glimpse C01}

\subsubsection{GC01 Narrow-Band Observations}

	The $(PAH, H_2O-PAH)$ and $(PAH, PAH-H3+)$ CMDs of GC01 are shown in 
Figure 9. The reddening vector has a near-vertical 
trajectory in both CMDs, and so differential reddening mainly blurs 
features in the CMDs along the magnitude axis. The smearing is 
expected to be $\pm 0.09$ magnitudes along the PAH axis based on 
the dispersion in the extinction found from the NIR TCD. Smearing along the 
color axis is modest, and the giant branch of GC01 is clearly seen in both CMDs.

\begin{figure}
\figurenum{9}
\epsscale{1.00}
\plotone{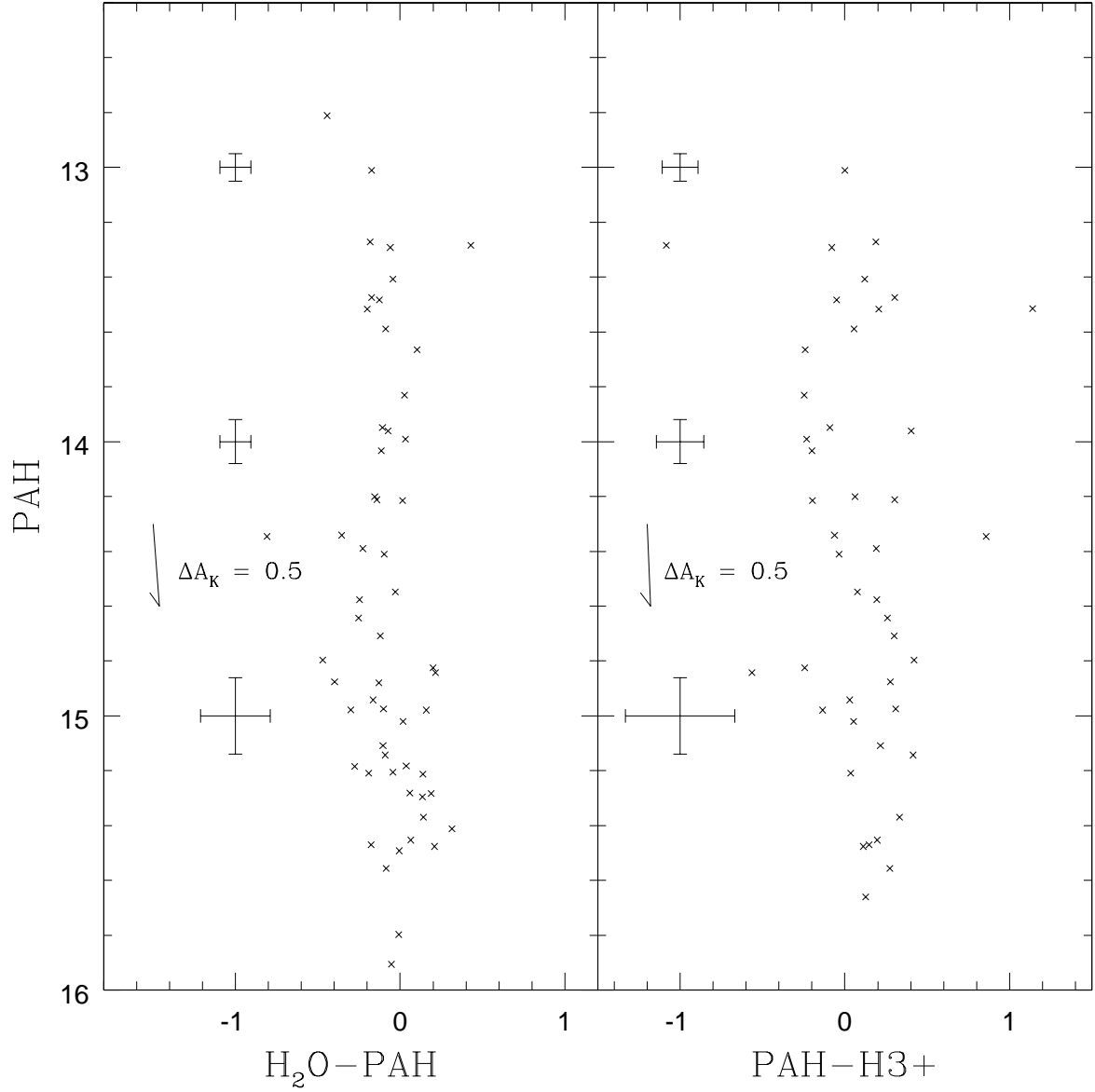}
\caption{$(PAH,H_2O-PAH)$ and $(PAH, PAH-H3+)$ CMDs of GC01. 
Magnitudes are in the AB system. The error bars show typical 
$\pm 1 \sigma$ uncertainties calculated by ALLSTAR. Reddening vectors that follow 
Equation 4 of Nishiyama et al. (2009) are also shown.} 
\end{figure}

	The narrow-band measurements can be used to check the reddening estimated 
from the TCD in the previous section. The mean SED of GC01 stars in the 
$1 - 3.5\mu$m interval, normalized to the signal in $K$, are shown in the top 
panel of Figure 10. Also plotted in Figure 10 is the SED of the K3III star 
HR8925 based on the $JHK$ magitudes and the flux density measurements given 
by Rayner et al. (2009). The SED of HR8925 has been reddened 
by applying the Nishiyama et al. (2009) reddening law for 
A$_K = 1.1$ and 1.4, which is the $\pm 1 \sigma$ range in A$_K$ found for 
GC01 in Section 3.1. The mean GC01 SEDs between 1 and $3.5\mu$m 
match that of the reddened HR8925 SED -- the long wavelength measurements are thus 
consistent with the extinction found at shorter wavelengths.

\begin{figure}
\figurenum{10}
\epsscale{1.00}
\plotone{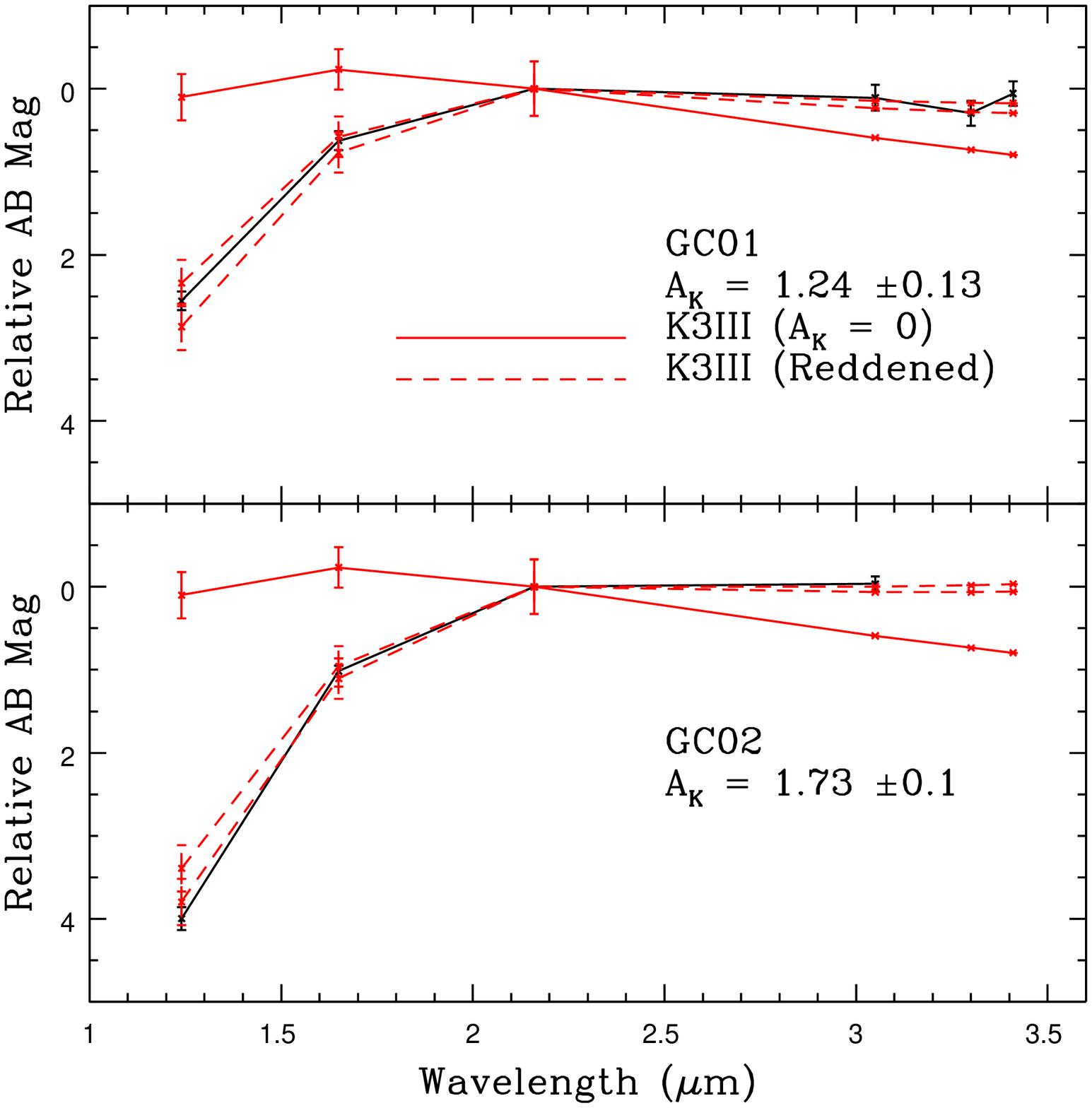}
\caption{The mean SEDs of bright stars in GC01 (top panel) and GC02 (bottom panel) 
in the $1 - 3.5\mu$m interval. All magnitudes are measured with respect to those 
in $K$. The error bars in the cluster measurements show the 
error in the mean at each wavelength. The solid red line is the SED of the K3III 
star HR8925, constructed from the magnitudes and flux densities given by Rayner et al. 
(2009). The error bars in the HR8925 curves are those cited by Rayner et al. 
(2009). The dashed lines show the SED of HR8925 after 
applying the Nishiyama et al. (2009) reddening law. There is good 
agreement between the cluster and HR8925 SEDs.}
\end{figure}

	The PAH observations of GC01 have an 
angular resolution that is $\sim 2 \times$ finer than the NIR 
measurements, and so can be used to assess crowding in the NIR data. The PAH 
LF constructed from sources in the $(PAH, H_2O-PAH)$ CMDs is shown in Figure 11. 
The onset of the PAH LF in Figure 11 occurs near $PAH \sim 13.5$.

\begin{figure}
\figurenum{11}
\epsscale{1.00}
\plotone{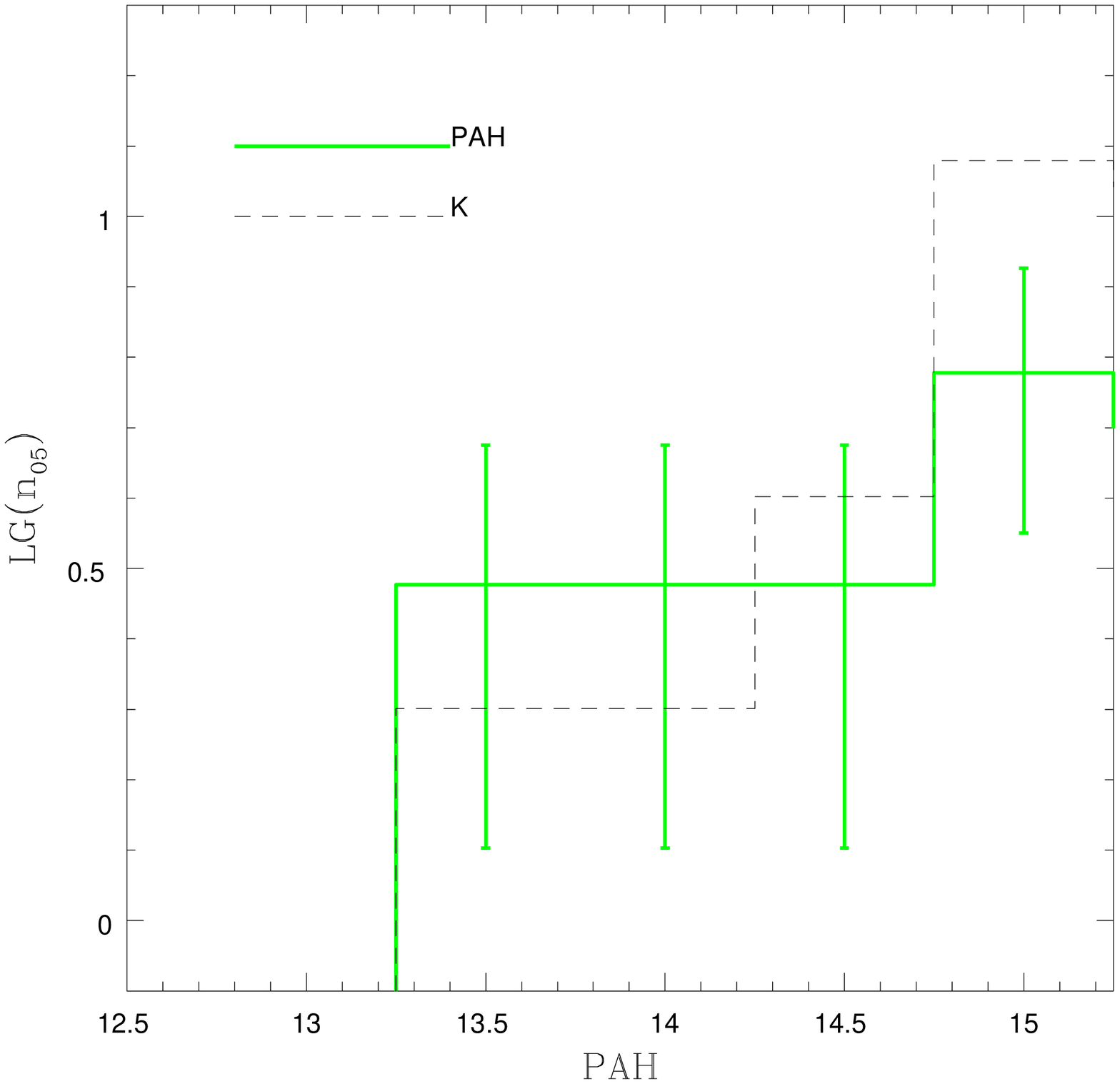}
\caption{LFs of GC01 in the PAH (solid green line) and 
$K$ (dashed black line) filters. The $K$ and PAH number counts are complete in 
the magnitude range shown, while the random photometric errors in this magnitude range 
are a fraction of the 0.5 magnitude bin size. The PAH number counts are taken from 
the (PAH, H$_2$O -- PAH) CMD, while those in $K$ are
from the $(K, H-K)$ CMD in Figure 2. The $K$ magnitudes have been shifted along 
the magnitude axis to account (1) for the $K-PAH$ color of a K giant and (2) for 
differences in extinction between the $K$ and PAH filters. Neither dataset 
has been corrected for differential reddening, and the LFs consider only 
sources that are in the area that is common to both the PAH and $K$ observations. 
That the LFs agree near the bright end indicates that blending is not 
a factor among the brightest stars in the $K$ image, despite the larger FWHM of 
the $K$ PSF.}
\end{figure}

	Also shown in Figure 11 is the $K$ LF of GC01, where only sources 
that are in the same area that was observed through the PAH filter have 
been counted. The PAH measurements have not been corrected for differential 
reddening, and so for consistency the $K$ number counts in Figure 11 were 
taken from the $(K, H-K)$ CMD, which also was not corrected for differential 
extinction. The $K$ LF in Figure 11 has also been shifted along the magnitude axis 
by an amount equal to the $K-PAH$ color of a K giant that is viewed through 
A$_K = 1.24$ magnitudes of extinction.

	If sources blend together they will appear as a single object 
that is brighter than the individual components. If the frequency of blending is high 
among the most luminous members of a system then a population of objects that is 
brighter than the individual brightest stars will be seen, and the overall effect of 
blending on the LF will be to shift it along the magnitude axis to brighter 
values. The two LFs in Figure 11 agree over a $\sim 2$ magnitude interval at 
the bright end, suggesting that crowding is not an issue among 
the brightest stars in the $K$ observations.

\subsubsection{GC01 SPITZER Observations}

	The $([4.5],[3.6]-[4.5])$ CMDs of sources in different annuli centered on GC01 
are shown in Figure 12. The vertical plume in $([4.5], [3.6]-[4.5])$ CMDs can be 
populated by a diverse mix of stars with a wide range of effective temperatures. 
Unlike at shorter wavelengths and with the exception of all but the coolest sources, 
there is only a small dispersion in the intrinsic 
[3.6]--[4.5] colors of stars as the [3.6] and [4.5] filters sample the descending 
red edge of the SED. 

\begin{figure}
\figurenum{12}
\epsscale{1.00}
\plotone{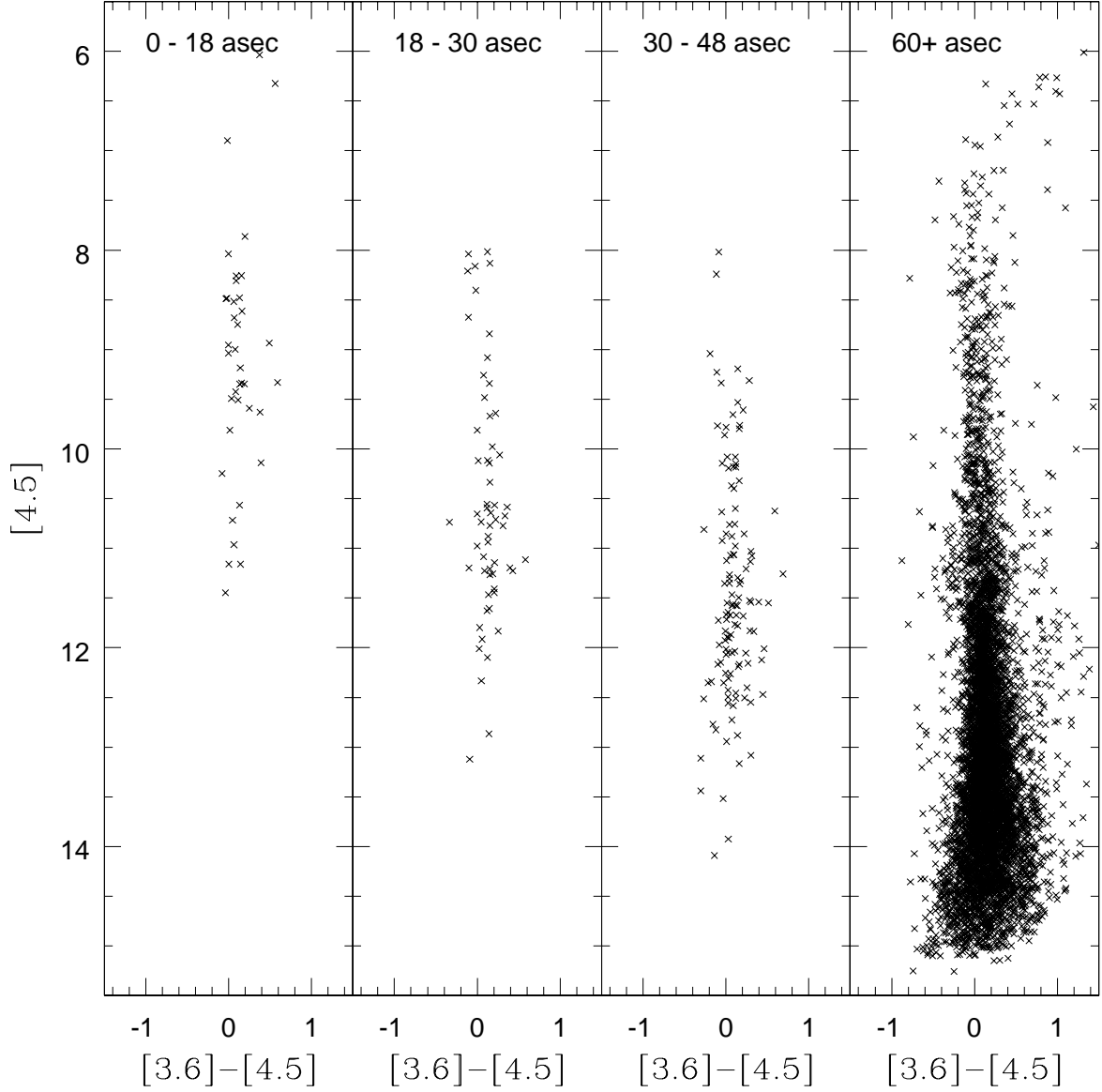}
\caption{$([4.5], [3.6]-[4.5])$ CMDs of GC01 and its surroundings. Number counts 
indicate that the CMD in the right hand panel is dominated by non-cluster stars, 
although a modest contribution from stars belonging to GC01 is 
present. Assuming a uniform distribution of field stars on the sky, 
number counts indicate that the CMDs that sample objects 
out to 48 arcsec from the cluster center are dominated by 
cluster objects. There is an intrinsic cut-off in the cluster 
sequence near [4.5]=8.0 in the middle two panels that we attribute to 
the RGB-tip.} 
\end{figure}

	Comparisons between star counts made from the SPITZER 
observations at various distances from the cluster center indicate that cluster 
members dominate the number counts out to 30 -- 48 arcsec from the center of GC01. 
Of the 80 stars with [4.5] between 8 and 12 in the 30 -- 48 
arcsec CMD in Figure 12, source counts in the $60+$ arcsec CMD suggest 
that only 15 of these are field stars if non-cluster stars are assumed 
to be uniformly distributed. The number of field stars is actually 
slightly lower than this as there is modest contamination from the outer 
regions of GC01 at radii $> 60$ arcsec (see the Appendix). 

	The $([4.5], [3.6]-[4.5])$ CMD of objects between 18 and 48 arcsec 
from the center of GC01 is compared with isochrones from Marigo 
et al. (2008) in Figure 13. A distance modulus of 13.6 and A$_K = 1.24$ has 
been assumed, with extinction applied according to the Nishiyama et al. (2009) 
reddening law. The 60\% silicate and 40\% AlOx mix for circumstellar dust 
from Groenewegen (2006) has been adopted, although the models are not sensitive 
to the chemistry of the circumstellar envelope at these wavelengths (e.g. Davidge 
2014). Models with solar and half-solar metallicities are shown.

\begin{figure}
\figurenum{13}
\epsscale{1.00}
\plotone{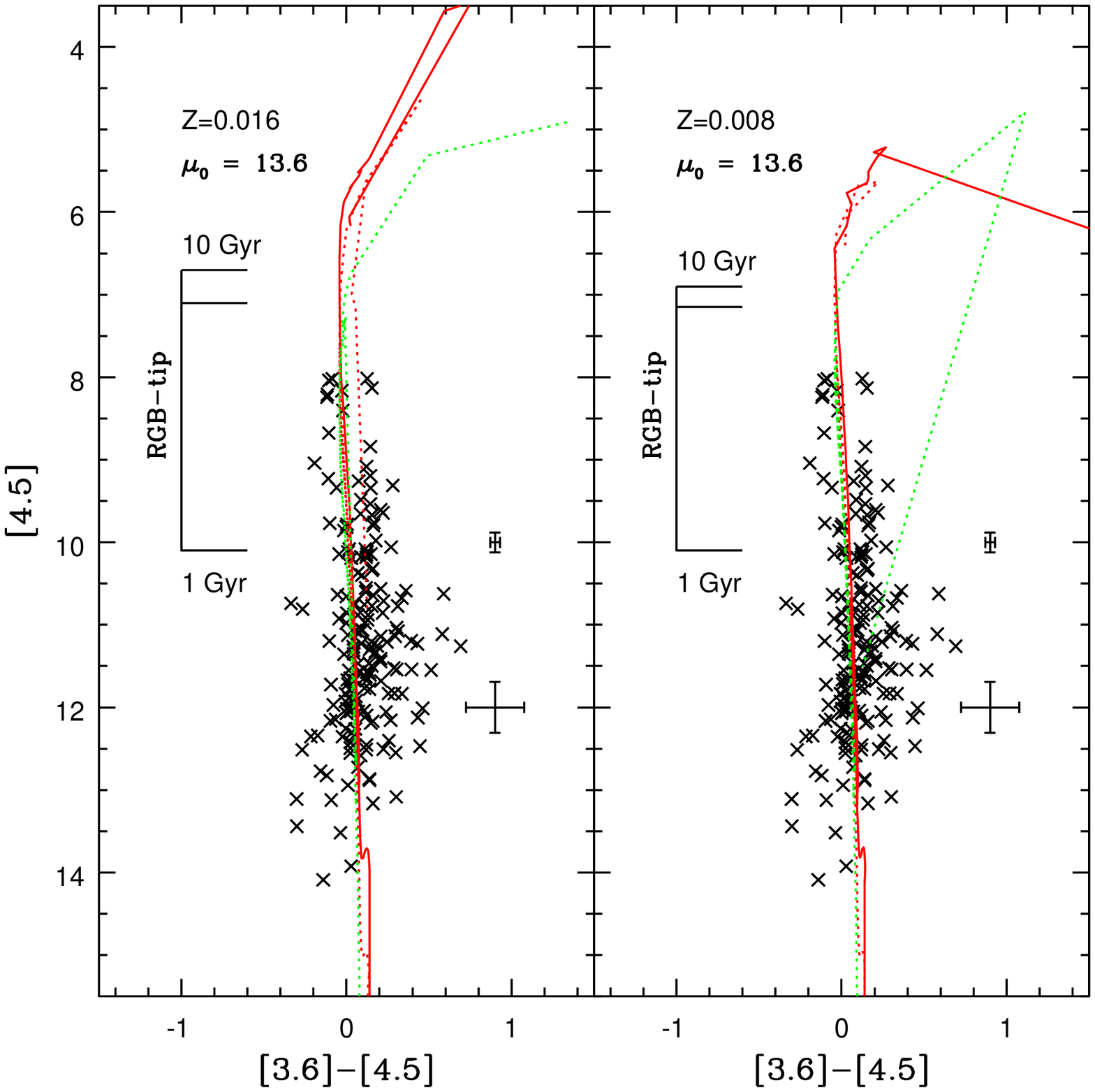}
\caption{Comparisons with solar (left hand panel) and half solar (right 
hand panel) metallicity isochrones from Marigo et al. (2008). The crosses show 
objects that are between 18 and 48 arcsec from the center of GC01, and the errorbars 
show $\pm 1 \sigma$ uncertainties from artificial star experiments. Models with ages of 
1 Gyr (solid red), 2.5 Gyr (dashed red), and 10 Gyr (dashed green) are shown for 
a distance modulus of 13.6 and A$_K = 1.24$. The predicted brightnesses of 
the RGB-tip for ages -- from bottom to top -- 1 Gyr, 2.5 Gyr, 
and 10 Gyr are also shown.}
\end{figure}

	Boyer et al. (2015) define extreme AGB stars to have unreddened 
[3.6]--[4.5] colors $ > 0.1$. If such objects are present in GC01 then they will form 
a population of objects with [4.5] $> 8$ that will also have [3.6]-[4.5] colors that 
exceed those of the GC01 locus as defined when [4.5] $> 8$. While we can not 
distinguish between AGB stars that do not have warm circumstellar dust envelopes and 
RGB stars based solely on [3.6]--[4.5] colors, cluster members that are brighter 
than the RGB-tip should be evolving on the AGB.

	The peak observed stellar brightness in a system depends 
not only on its age and metallicity, but also on the overall mass of the system, 
as there is a low probability of occupation in the portions of CMDs that sample rapid 
phases of evolution. The isochrones predict that the AGB may extend to [4.8] $\leq 5$ 
at the distance of GC01. There are objects as bright as [4.5] = 6 in the innermost 
annulus. Given that all stars in the middle two panels 
have [4.5] $\geq 8$ then if the stars with [4.5] = 6 are blends they must be 
unresolved asterisms that are made up of multiple stars. Such blending is feasible 
given the density of moderately bright objects in our field (e.g. Figure 11).

	The CMDs of objects located between 18 and 48 arcsec from the cluster 
center indicates that there is a drop in star counts when [4.5] $\leq  
8$. This is not due to saturation in the cores of stellar images, 
as numerous stars that have [4.5] between 8 and 6 magnitudes are seen in the 
right hand panel of Figure 12. We note that [4.5] = 8 is more-or-less consistent with 
the peak brightness found by Ivanov et al. (2005) in $K$ if it is assumed that the 
stars do not have excess thermal emission, as expected given their [3.6]--[4.5] 
colors. The expected location of the RGB-tip is indicated for 
each model, and -- unless GC01 has an age $< 1$ Gyr -- the vast majority of stars 
detected in the SPITZER images within 60 arcsec of the center of GC01 are evolving 
on the RGB. To the extent that the distance modulus and models are correct and that 
the RGB-tip brightness occurs near [4.5] = 8 then 
the models suggest that GC01 has an age between 1 and 2.5 Gyr. 

	The [4.5] LF of GC01 stars in the 18 -- 48 arcsec interval is shown in 
Figure 14. The LF is restricted to [4.5] $< 11.5$, 
as this is where the artificial star experiments suggest that sample 
completeness exceeds 50\%. A statistical correction for non-cluster stars 
was made by subtracting number counts of sources that are more than 60 
arcsec from the cluster center after scaling to account for differences in 
area. While diffuse cluster light can be traced to radii in excess of 60 
arcsec (see the Appendix), the density of cluster stars at these radii 
is much lower than in the inner regions of the cluster, and the star counts are 
dominated by field stars. In fact, the correction for field stars produces only a 
modest change in number counts in the GC01 LF. Artificial star experiments indicate 
that the uncertainties in [4.5] are $\leq \pm 0.1$ magnitude in the magnitude 
range shown, and so bin-to-bin blurring is modest.

\begin{figure}
\figurenum{14}
\epsscale{0.90}
\plotone{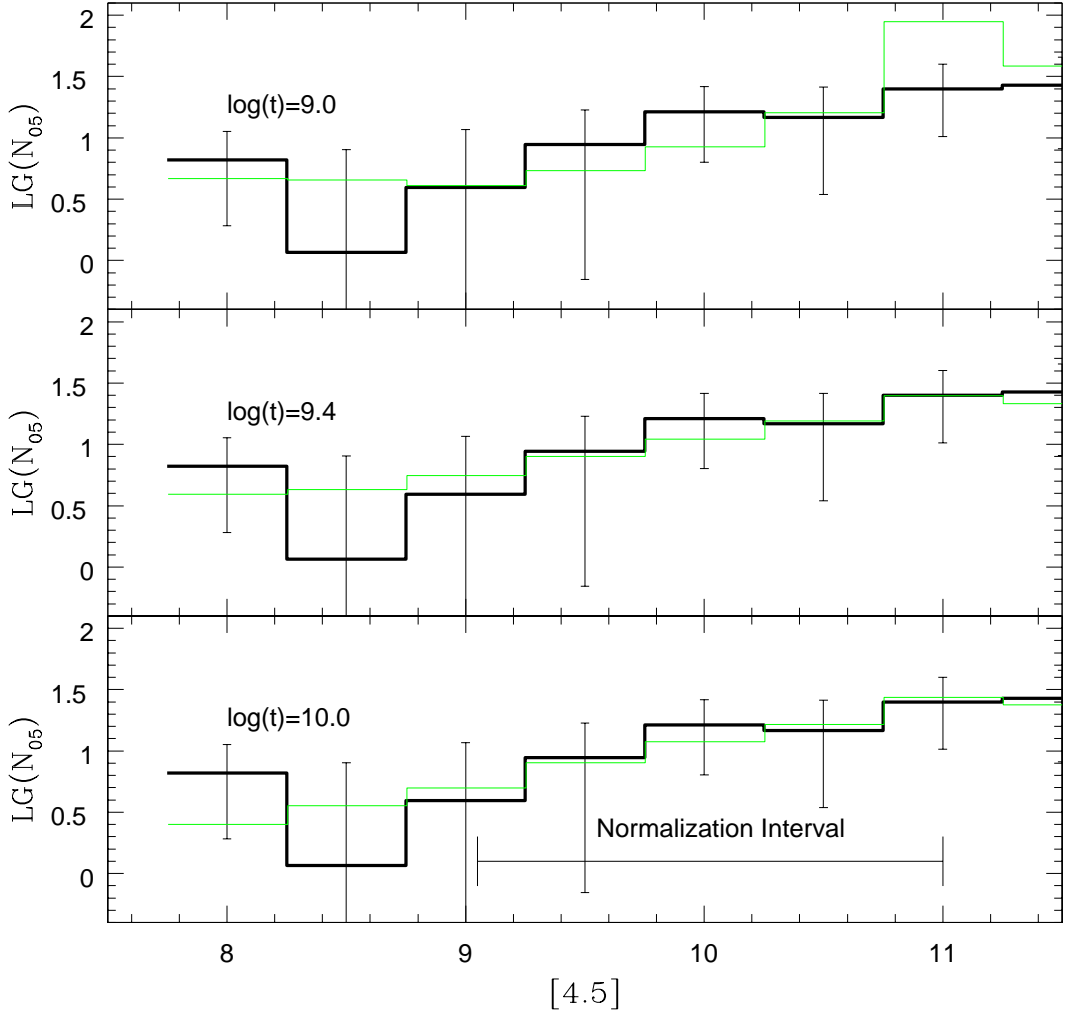}
\caption{[4.5] LF of GC01, based on the $([4.5], [3.6]-[4.5])$ 
CMDs of objects between 18 and 48 arcsec from the cluster center. 
The error bars show $\pm 1 \sigma$ uncertainties computed from counting 
statistics. A correction for non-cluster members has been made 
by subtracting the area-scaled LF of sources with $r > 60$ arcsec. 
Model LFs from solar metallicity Marigo et al. (2008) isochrones are also shown. 
A distance modulus of 13.6 and $<A_K> = 1.24$ have 
been assumed, with the models shifted along the vertical axis to match the number 
counts in the magnitude interval that is indicated. The 
models match the slope of the GC01 LF between [4.8] = 8 and 11.}
\end{figure}

	Comparisons are made with model LFs in Figure 14. 
The model LFs do not vary greatly with metallicity in the range of magnitudes 
considered, and so only solar metallicity models are shown. The models have 
been shifted along the vertical axis to match the number counts between [4.5] = 9 
and 11 to avoid magnitudes where the sample is not complete. The overall 
shape of the LF is consistent with that predicted by models of stellar 
evolution. This is a robust result, that is largely independent of the assumed age.

\subsection{Glimpse C02}

\subsubsection{GC02 Narrow-Band Observations}

	Given that only a modest number of stars were detected in the H$_2$O 
observations of GC02, we do not examine the $(H_{2}O, K-H_{2}O)$ 
CMD of this cluster. The mean SED of bright stars near the center of GC02 
in the $1 - 3.1\mu$m interval is examined in the lower panel of Figure 10. 
It should be recalled that GC02 was not observed through the PAH and H3+ filters 
because of the inherent faintness of the member stars (Section 2). 
The lack of PAH and H3+ measurements notwithstanding, the 
H$_2$O observations extend the mean SED well past $2\mu$m. The dashed red lines 
show the SED of the K3III star HR8925 reddened by $\pm 0.1$ magnitude about the A$_K$ 
found from the $(K, H-K)$ CMD. The SEDs of bright stars 
observed near the center of GC02 are consistent with them being highly 
reddened late-type giants. 

\subsubsection{GC02 SPITZER Observations}

	The ([4.5], [3.6]-[4.5]) CMDs of objects in four radial intervals 
centered on GC02 are shown in Figure 15. A vertical plume of 
cluster members is evident at small radii. There are no objects 
near the top of the GC02 CMD with [3.6]--[4.5] colors that fall redward of the 
main locus of points, which would be candidate highly evolved AGB 
stars belonging to GC02. The brightest stars in the 24 -- 36 arcsec 
CMD have magnitudes that are comparable to those in the 0 -- 18 arcsec CMD, 
suggesting that the brightest objects detected near the center of GC02 in 
the [3.6] and [4.5] images may not be blends.

\begin{figure}
\figurenum{15}
\epsscale{1.00}
\plotone{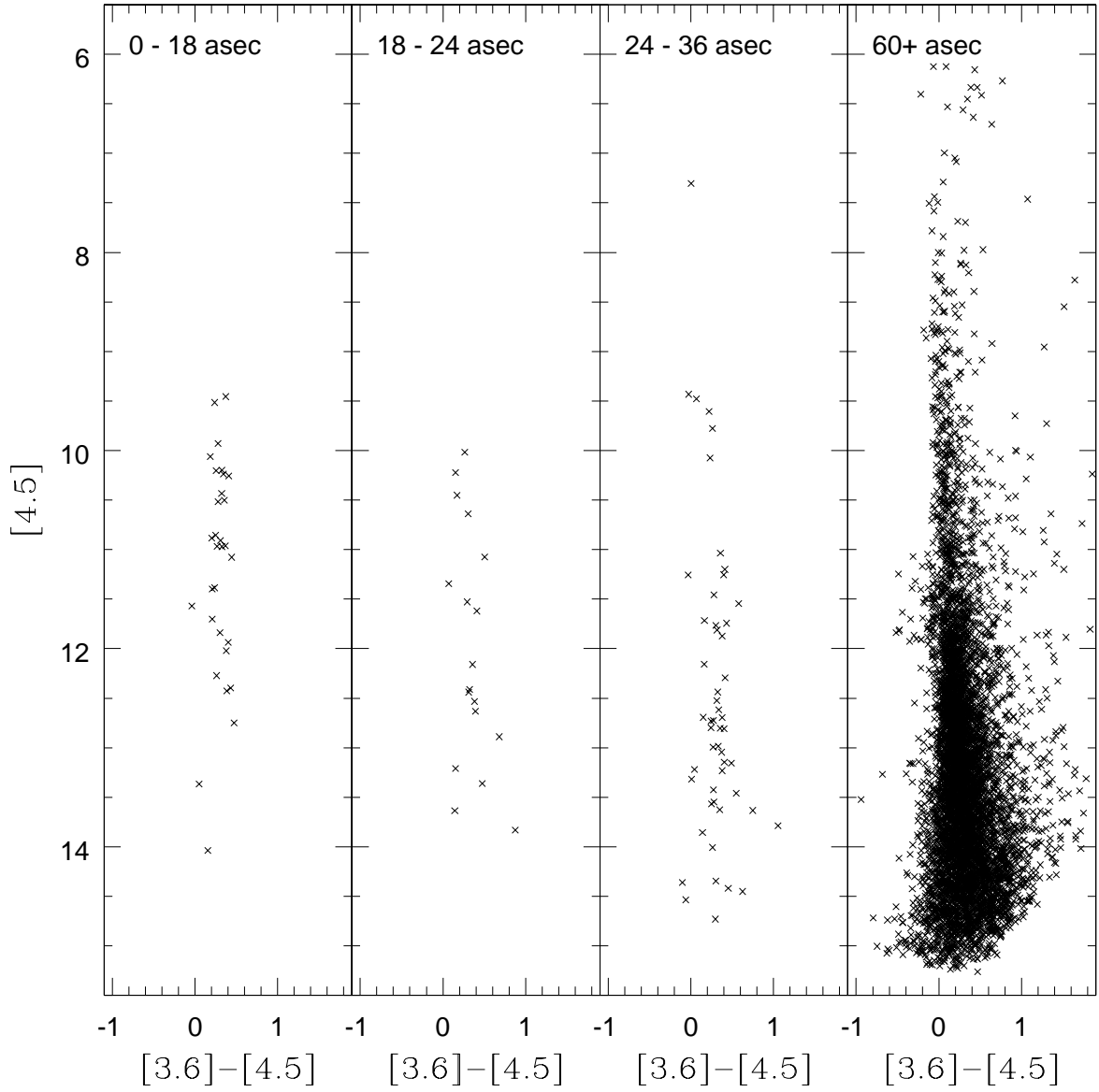}
\caption{Same as Figure 12, but showing $([4.5], [3.6]-[4.5])$ CMDs of GC02.}
\end{figure}

	The CMD of GC02 is compared with isochrones from Marigo et al. (2008) in 
Figure 16. There is a $\sim 0.2$ magnitude offset along the [3.6]--[4.5] axis 
between the GC02 sequence and the models.  A similar offset is not seen in 
the GC01 photometry (e.g. Figure 13). We have compared our PSF-based photometric 
measurements for a sample of isolated objects with [3.6] between 9 and 10
with those in published GLIMPSE source catalogs made from aperture measurements, and 
find agreement to within a few hundredths of a magnitude. Thus, the offset 
in [3.6]--[4.5] is not the result of errors in our photometric measurements.

\begin{figure}
\figurenum{16}
\epsscale{1.00}
\plotone{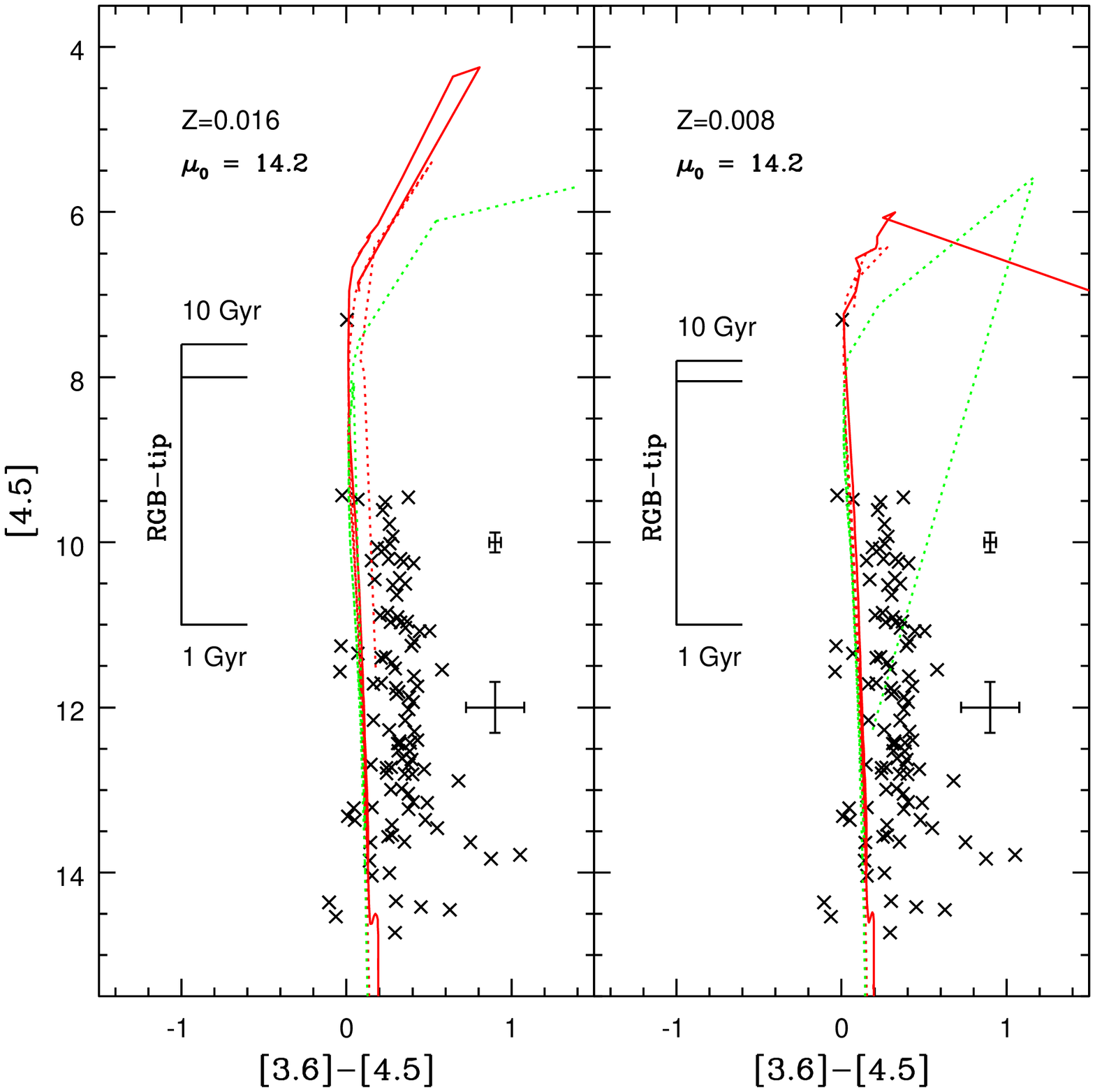}
\caption{Same as Figure 13, but showing the CMD of sources 
within 36 arcsec of the center of GC02. The isochrones assume A$_K = 1.73$
with a distance modulus 14.2 (Section 3.2). The models predict that the vast 
majority of stars in the CMD are evolving on the RGB. The object with 
[4.5] = 7.3 is likely evolving on the AGB if it is a cluster member. The 0.2 magnitude 
offset in [3.6]--[4.5] between the observations and models is discussed in the text.}
\end{figure}

	We are unsure as to the origin of the offset in [3.6]--[4.5] color, although 
there is a tendency in the $r > 60$ arcsec CMD for objects with [4.5] $> 12$ to have 
larger [3.6]--[4.5] colors than those with [4.5] $< 12$. There is heavy extinction 
towards GC02, and a correlation between magnitude and color will occur if the brightest 
stars, many of which are presumably nearby if they are not cluster members, are subject 
to lower levels of extinction than the fainter objects, which presumably tend to be 
more distant, and so have a greater chance of being more heavily obscured. 
However, the size of the offset in [3.6]--[4.5] is hard 
to explain with reddening. Uncertainties in A$_K$ of a few magnitudes have only a minor 
impact on the position of the models at these wavelengths, and the extinction towards 
GC02 would have to be A$_K \sim 4$ to produce the color difference. As for the 
possibility of an abnormal reddening law towards GC02, 
the variations in line of sight extinction that are seen at visible wavelengths 
are much reduced at wavelengths longward of $1\mu$m (e.g. Indebetouw et al. 2005). 
Given the unexplained red [3.6]--[4.5] colors we caution that the [3.6] and [4.5] 
photometry of GC02 may have uncertainties of a few tenths of a magnitude.

	Uncertainties in the photometry on the scale of 10 - 20\% notwithstanding, 
the models predict that stars evolving on the AGB in GC02 will depart 
from a near-vertical trend $\sim 2 - 3$ magnitudes above [4.5]=9.5, which is the 
brightness that we assign to the RGB-tip. The uncertainties 
in the calibration of the SPITZER data near GC02 
indicate that the RGB-tip brightness is uncertain by a few tenths of a magnitude. 
Still, this [4.5] magnitude for the RGB-tip corresponds to $K \sim 10$, which 
is the magnitude of the brightest star along the cluster ridgeline drawn in the middle 
panel of Figure 2 of Kurtev et al. (2008). The isochrones 
suggest that the majority of stars detected in the SPITZER images are evolving on the 
RGB, although there is one object with [4.5] = 7.3 that may be on the AGB 
if it is a cluster member. The intrinsic brightness of the RGB-tip in GC02 is 
similar to that in GC01, and if the RGB-tip occurs near [4.5]=9.5 then the 
isochrones predict an age between 1 and 2.5 Gyr. 

	The [4.5] LF of stars between 0 and 36 arcsec in GC02 is shown in 
Figure 17. The entries have been corrected for non-cluster sources by subtracting 
the LF of objects with $r > 60$ arcsec after adjusting for differences in 
areal coverage. As was the case for GC01, while stars that belong to GC02 are 
present at radii $> 60$ arcsec, their number density is low when compared with those at 
smaller radii (e.g. the Appendix), and stars in the field dominate the number counts. 
The fractional contamination by non-cluster stars becomes 
significant when [4.5] $> 12$, and so only this part of the LF is shown.

\begin{figure}
\figurenum{17}
\epsscale{1.00}
\plotone{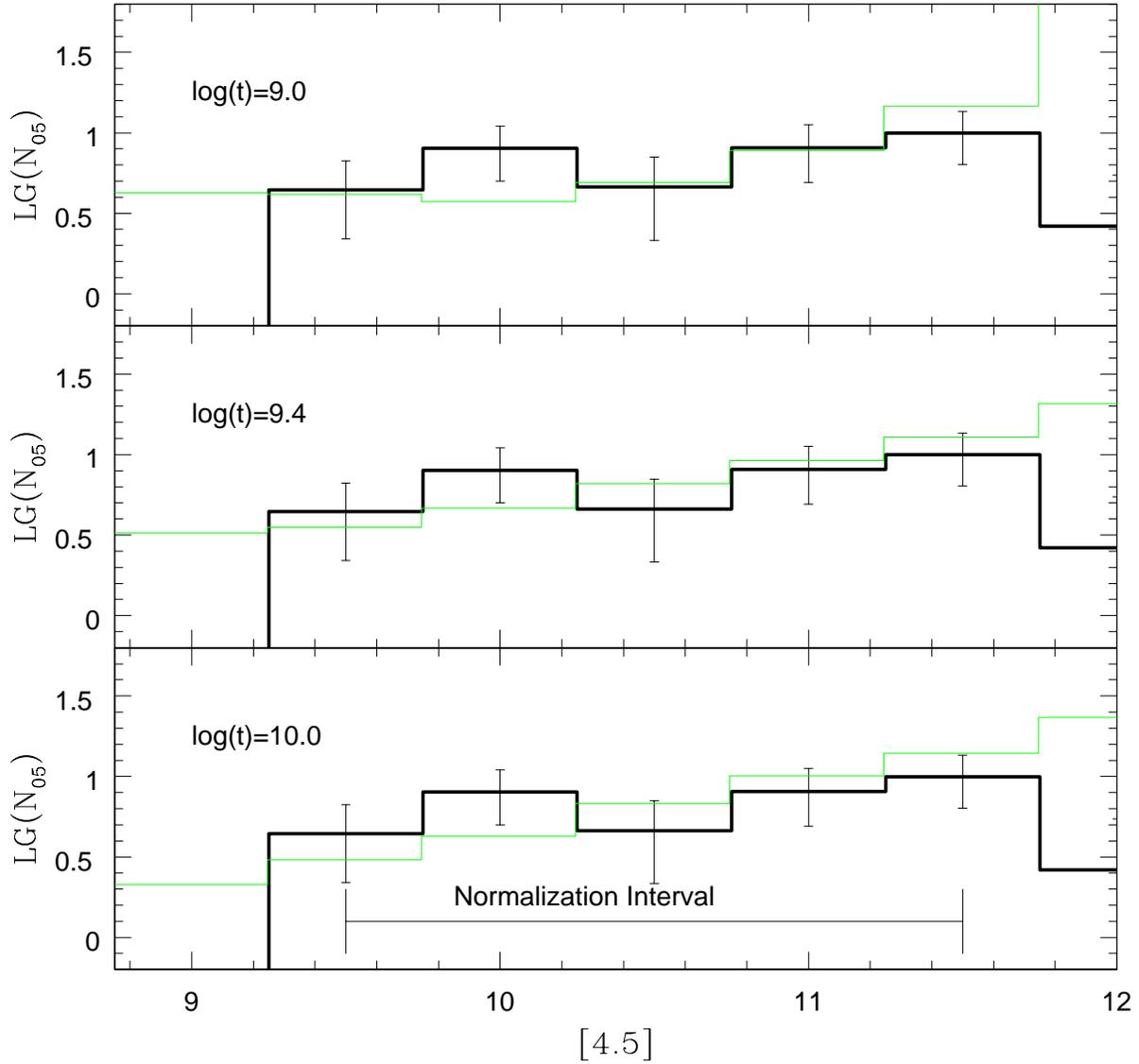}
\caption{The [4.5] LF of GC02. A statistical correction for non-cluster 
members was applied by subtracting a scaled version of the LF of sources at $r > 60$ 
arcsec from the cluster center. This correction does not affect the shape of the GC02 
LF when [4.5] $< 12$. Model LFs constructed from solar metallicity Padova isochrones 
are shown for a distance modulus of 13.7 and A$_K = 1.7$.}
\end{figure}

	The [4.5] LF is compared with solar metallicity model LFs in Figure 17. 
The statistical significance of any difference between the observations 
and the various models in Figure 17 is low. The shallow nature of the GLIMPSE 
survey prevents the RC from being sampled in GC02, and this severly limits 
conclusions that might otherwise be drawn from comparisons with model LFs. 

\section{DISCUSSION \& SUMMARY}

	AO-corrected images that span the $1 - 3.5\mu$m wavelength interval 
have been used to probe the stellar contents of the star clusters GC01 
and GC02. These clusters are heavily reddened and are subject to significant 
contamination from non-cluster stars owing to their location at low 
Galactic latitudes. NIR and MIR imaging of their central regions, 
where the fractional contamination from foreground 
and background stars is lowest, offers a promising means 
of determining their age and distance. 

	$JHK$ and narrow-band images, with the latter sampling the $3 - 3.5\mu$m 
wavelength interval, were recorded with the IRCS and the RAVEN AO 
science demonstrator on the Subaru telescope. Stars in the narrow-band images have 
a FWHM that is within a few hundredths of an arcsec of the telescope diffraction 
limit, demonstrating that good image quality can 
be delivered by MOAO systems that work in open-loop. 
While the narrow-band images are shallower than the NIR images, they 
provide a means of checking if crowding has affected photometric measurements 
obtained from images that have poorer angular resolutions. The narrow-band 
images also allow the SEDs of cluster stars to be extended into the MIR, providing 
additional wavelength leverage for checking reddening estimates. 
The SEDs obtained here cover the $1 - 3.5\mu$m wavelength interval, and are 
consistent with the brightest objects in each field being heavily reddened K 
giants. The combined NIR and MIR SEDs are consistent with the mean reddenings obtained 
from the NIR photometry, which are A$_K = 1.24 \pm 0.13$ for GC01 and A$_K = 1.73 
\pm 0.10$ for GC02.

	Archival [3.6] and [4.5] Spitzer images that were recorded for the 
GLIMPSE survey have also been examined. 
While having an angular resolution that is almost an order of 
magnitude larger than that of the IRCS $+$ RAVEN images, 
the angular coverage of the GLIMPSE survey 
allows a statistical assessment of non-cluster sources to be made. 
The CMDs constructed from the SPITZER data 
provide information about the luminosity and spatial 
distribution of bright stars in each cluster. The RGB-tip measurements obtained 
from the SPITZER data are consistent with those made previously in the NIR. 

	We preface our discussion of these clusters with a cautionary note. 
The bright stellar content in the central regions of some dynamically evolved 
clusters is not representative of what is seen outside of their cores 
(e.g. Davidge 1995), and this may bias efforts to probe stellar content.
Number counts made from the SPITZER images suggest that stars 
in GC01 can be detected in statistically significant 
numbers with respect to foreground/background objects out to at least 30 arcsec 
from the cluster center, and the same holds for GC02. Assuming a distance of 5.2 kpc 
for GC01, then 30 arcsec corresponds to $\sim 0.8$ parsec, which is 
comparable to the core radius of a typical globular cluster, and is 
an order of magnitude smaller than the typical half light radius 
(e.g. van den Bergh \& Mackey 2004). In the Appendix we 
show that light from GC01 and GC02 can be traced out to radii of at least 
100 arcsec. Thus, a rich population of cluster members awaits discovery 
in the large areas of GC01 and GC02 that have not been explored to date, and these 
stars will undoubtedly provide additional clues about the distance and age 
of these clusters. Contamination from non-cluster stars presents a daunting obstacle 
for efforts to identify cluster stars at large radii.

\subsection{GC01}

	GC01 is of interest for studies of the evolution of the Galactic disk 
because it is one of the most massive clusters that may have formed during intermediate 
epochs (Davies et al. 2011). The formation of large, compact clusters is often 
associated with interactions and/or starburst events (e.g. Ashman \& Zepf 2001). 
Does the age of GC01 coincide with a past event that may have influenced Galactic 
evolution? Davies et al. (2011) conclude that GC01 is not an old globular cluster, 
and assign it an age between 0.3 and 2 Gyr, with the most probable age between 
0.4 and 0.8 Gyr. Davies et al. (2011) further suggest that the formation of GC01 
may be linked to a past encounter between the Galactic 
disk and the Magellanic Clouds. In fact, Rezaei kh. et al. (2014) find peaks 
in the SFRs of the LMC and SMC $\sim 0.7$ Gyr in the past, which they suggest 
may be linked to an interaction with the Galaxy.

	The observations discussed here do not support the 
formation of GC01 within the past Gyr, although the amplitude of the RC in the 
$K$ LF is consistent with the older end of the Davies et 
al. (2011) age range. While our data suggest that GC01 may 
be too old to have formed as part of the most recent interaction 
with the Magellanic Clouds, it does not rule out its formation 
during previous interactions. This being said, proper motion measurements 
suggest that the Magellanic Clouds may either be on their first approach to the Galaxy 
or that their orbital period about the Galaxy is much longer than once 
thought (Besla et al. 2007). In any event, the presence of young clusters with 
masses $> 10^4$M$_{\odot}$ like Westerlund 1 and the Arches suggests that clusters with 
masses approaching that of GC01 may form naturally 
throughout the lifetime of the Galaxy, without the need of an external trigger. 

	Uncertainties in the origins of GC01 notwithstanding, its study 
may provide clues into the evolution of compact intermediate age clusters. 
There are hints that star formation is occuring along the GC01 line of sight. 
SPITZER images of GC01 show a prominent dust lane 
projected against the cluster (Kobulnicky et al. 2005), and the star-to-star 
differences in mean extinction found in Section 3 indicate that the dust 
distribution is clumpy, with a characteristic size that is consistent with 
that of individual stellar systems (e.g. Larson 1995). There are also 
candidate young stellar objects (YSOs) seen near GC01 on the sky 
(Kobulnicky et al. 2005). It is not known if the dust clumps and candidate YSOs 
are physically associated with the cluster, or are chance superpositions.

	Goudfrooij et al. (2014) present evidence for multiple periods of star-forming 
activity in massive LMC clusters, and investigate the characteristics of clusters 
where such activity might occur. Mechanisms other than multiple episodes 
of star formation have been proposed to explain the properties of these 
clusters (e.g. Brandt \& Huang 2015; Niederhofer et al. 2015, and references therein). 
The estimated mass of GC01 falls within the range of LMC clusters 
where it has been suggested that multiple episodes of star formation 
have occured, and it is intriguing that the $K$ LF of GC01 shows 
characteristics at the bright end that are consistent with young and intermediate 
age populations, while the faint end of the LF is more consistent with that of an old 
population.

	The age estimate gleaned here from evolved stars can be checked 
by measuring the brightness of the MSTO. However, differential reddening 
smears the photometric measurements, thereby complicating this task. 
One strategy to reduce the impact of differential reddening would be to use 
deep AO-corrected integral field unit spectroscopy in the NIR to 
identify candidate MSTO stars. If spectral types can be 
established then intrinsic colors can be assigned, making it 
possible to construct a de-reddened CMD that samples the MSTO.

	While GC01 is viewed through A$_V \sim 11$ magnitudes of extinction, spectra of 
its integrated light at optical wavelengths will also provide insights into its 
stellar content and metallicity. The detection of deep Balmer 
absorption lines would be one signature of an intermediate age population.
The metallicity of GC01 could also be measured from the strengths 
of various atomic and molecular features in the integrated spectrum. 
Given that GC01 falls within the Solar Circle then 
a solar or supersolar metallicity would argue 
that it formed {\it in situ}. A metallicity that is one-half solar would be 
consistent with it having formed from material that likely originated outside of 
the Solar Circle if its age is less than a few Gyr.

\subsection{GC02}

	GC02 is a challenging target for stellar content studies as it is 
heavily extincted, although the absorbing material appears to be uniformly distributed. 
Kurtev et al. (2008) consider GC02 to be an old, metal-rich globular cluster. 
However, the the $K$ LF constructed from the RAVEN$+$IRCS observations 
does not show the onset of the SGB that is expected if the cluster is old. 
Tighter constraints on the age of GC02 could be obtained using 
deeper, diffraction-limited NIR images. For example, 
if GC02 is old but is more distant than assumed here then the SGB should 
show up with deeper images. If GC02 has an intermediate age 
and there is no large-scale differential reddening then it 
should be possible to detect the MSTO in the $(K, J-K)$ CMD 
of GC02. Given the distance and mean reddening of GC02 then the MSTO 
should occur near $K = 18 - 19$ and $J = 23 - 24$ if stars as young as 
1 Gyr are present. While the $(K, H-K)$ CMD of GC02 in Figure 6 does reach 
the required depth in $K$, the expected separation in 
color between the giant branch and 1 Gyr main sequence stars on the 
$(K, H-K)$ CMD at this brightness is only $\sim 0.2$ magnitudes, which is 
comparable to the uncertainties in the photometry at this magnitude. 
The integrated spectrum of GC02 at visible - red wavelengths 
should also contain deep Hydrogen lines if it has an age $\sim 1 - 2$ Gyr.

\acknowledgements{Sincere thanks are extended to Dr.Tae-Soo Pyo of Subaru Telescope,
who kindly supported our operation of the IRCS. The anonymous referee also provided 
numerous comments that helped improve the paper.}

\appendix

\section{The Light Profiles of GC01 and GC02}

	Published light profiles of GC01 (Kobulnicky et al. 2005; Ivanov et al. 2005; 
Davies et al. 2011) and GC02 (Kurtev et al. 2008) are based on star counts and/or 
isophotal measurements. These are restricted to the central 
few tens of arcsec in each cluster, as this is where the number density of cluster 
members exceeds that of field stars. The light profiles of GC01 and GC02 can be 
extracted out to much larger radii if the brightest resolved field stars are suppressed 
and the cluster light profiles are azimuthally smoothed to boost the 
signal-to-noise ratio. We demonstrate this using GLIMPSE [3.6] images.

	The [3.6] image of each cluster was rotated about the cluster center in 
15$^o$ increments and the rotated images were then combined by taking the median flux 
at each rotated pixel location. This process multiplexes the faint signal from the 
cluster light profiles at large radii and suppresses signal from individual objects, 
the majority of which will be field stars at large radii. Some artifacts of individual 
stars survive the median combination procedure, and these were suppressed 
by applying a $5.1 \times 5.1$ arcsec (i.e. $3\times$ the FWHM 
of the [3.6] PSF) running top-hat filter. This procedure 
tacitly assumes circular isophotes, and obliterates information over the 
angular scale of the smoothing filter, with the result that the light 
distribution near the cluster center can not be tracked.

	The [3.6] surface brightness profile of each cluster is shown 
in Figure A1, and light can be traced out to distances of at least 100 arcsec from 
the cluster centers. GC01 is much more centrally concentrated than GC02, 
and the light profile of the former may be truncated at radii $\sim 110$ 
arcsec. Signal from GC02 can be traced out to at least 200 arcsec.

\begin{figure}
\figurenum{A1}
\epsscale{1.00}
\plotone{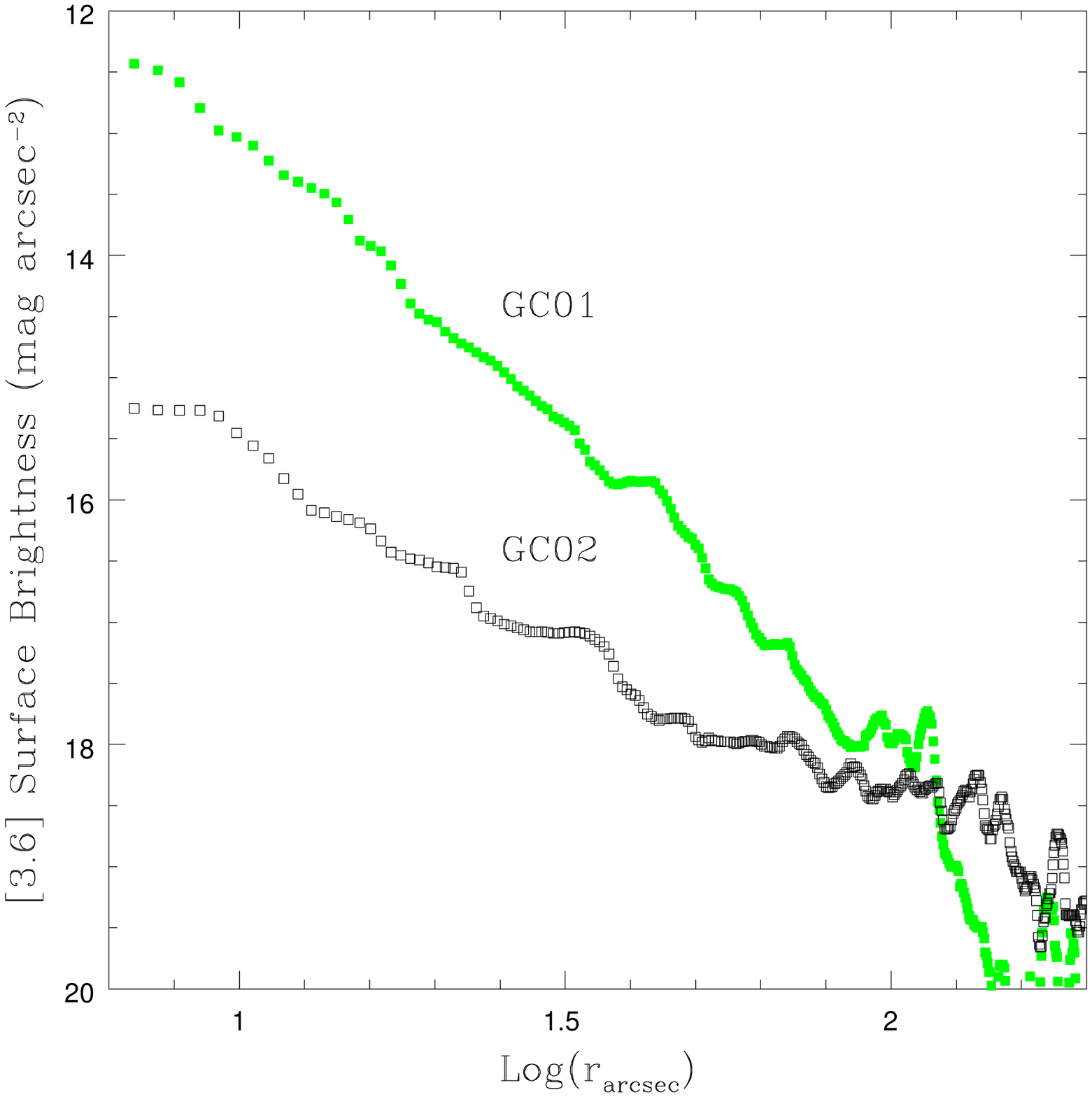}
\caption{[3.6] surface brightness profiles of GC01 and GC02, 
constructed using the azimuthal median-filtering technique described in the 
text. A $5.1 \times 5.1$ arcsec top-hat filter was applied to suppress 
artifacts of incomplete field star removal. Information on angular scales 
$< 6$ arcsec is thus lost, and so the lower radial limit for this 
plot is set at 6 arcsec. Light from both clusters can be traced out to distances 
$> 100$ arcsec from the cluster centers.}
\end{figure}

\end{document}